\documentclass{aa}

\usepackage{graphicx}
\usepackage{natbib}
\usepackage{scalerel}
\usepackage{amsmath}	
\usepackage{multicol}        
\usepackage{bm}		
\usepackage{pdflscape}	
\usepackage[usenames,dvipsnames,table]{xcolor}
\usepackage{tabularx}
\usepackage{multirow}
\usepackage{multicol}

\newcommand{\ATLAS}{$\textrm{ATLAS}^{\textrm{3D}}$}

\usepackage{txfonts}
\usepackage[pdfencoding=auto,psdextra]{hyperref}
\hypersetup{
    colorlinks=true,
    linkcolor=blue,
    filecolor=magenta,      
    urlcolor=blue,
    citecolor=blue
}
\makeatletter
\renewcommand*\aa@pageof{, page \thepage{} of \pageref*{LastPage}}
\makeatother

\usepackage[utf8]{inputenc}

\usepackage[switch, modulo]{lineno}
\linenumbers

\begin{document}
\nolinenumbers
%
%

\title{CASCO: Cosmological and AStrophysical parameters from Cosmological simulations and Observations}
\subtitle{II. Constraining cosmology and astrophysical processes with early- and late-type galaxies}

   
\newcommand{\orcid}[1]{} 

\author{V. Busillo,$^{1,2,3}$\thanks{E-mail: valerio.busillo@inaf.it}
C. Tortora,$^{2}$\thanks{E-mail: crescenzo.tortora@inaf.it}
G. Covone,$^{1,2,3}$
L. V. E. Koopmans,$^{4}$
M. Silvestrini,$^{1,2}$
N. R. Napolitano.$^{1}$}

\institute{$^{1}$Dipartimento di Fisica “E. Pancini”, Universit\`{a} degli studi di Napoli Federico II, Compl. Univ. di Monte S. Angelo, Via Cintia, I-80126 Napoli, Italy\\
$^{2}$INAF -- Osservatorio Astronomico di Capodimonte, Salita Moiariello 16, I-80131, Napoli, Italy\\
$^{3}$INFN, Sez. di Napoli, Compl. Univ. di Monte S. Angelo, Via Cintia, I-80126 Napoli, Italy\\
$^{4}$Kapteyn Astronomical Institute, University of Groningen, P.O Box 800, 9700 AV Groningen, The Netherlands\\}

%
%
\abstract{
Physical processes can influence the formation and evolution of galaxies in diverse ways. It is essential to validate their incorporation into cosmological simulations by testing them against real data encompassing various types of galaxies and spanning a broad spectrum of masses and galaxy properties. For these reasons, in this second paper of the CASCO series, we compare the structural properties and dark matter content of early-type galaxies taken from the \textsc{camels} IllustrisTNG cosmological simulations to three different observational datasets (SPIDER, \ATLAS\ and MaNGA DynPop), to constrain the value of cosmological and astrophysical feedback parameters, and we compare the results with those obtained comparing the simulation expectations with late-type galaxies. We consider the size-, internal DM fraction- and internal DM mass-stellar mass relations for all the simulations, and search for the best-fit simulation for each set of observations. For SPIDER, we find values for the cosmological parameters in line with both the literature and the results obtained from the comparison between simulations and late-type galaxies; results for the supernovae feedback parameters are instead opposite with respect to the previous results based on late-type galaxies. For \ATLAS, we find similar values as from SPIDER for the cosmological parameters, but we find values for the supernovae feedback parameters more in line with what we found for late-type galaxies. From MaNGA DynPop, we find extreme values for the cosmological parameters, while the supernovae feedback parameters are consistent with \ATLAS\ results. When considering the full MaNGA DynPop sample, including both late- and early-type galaxies, no single simulation can reproduce the full variety in the observational datasets. The constraints depend strongly on the specific properties of each observational trend, making it difficult to find a simulation matching all galaxy types, indicating the existence of limitations in the ability of simulations in reproducing the observations.
}
%
%
%
%
   \titlerunning{CASCO-II}
   \authorrunning{V. Busillo}
   
   \maketitle
%
%
%
\section{Introduction}

The study of galaxy formation and evolution is a cornerstone of modern astrophysics. One of the key challenges in this field is to understand the complex interplay between cosmology and astrophysical feedback mechanisms, and how both shape the observable properties of galaxies.

In this regard, cosmological simulations constitute a useful tool to investigate this interplay. For example, the SEAGLE programme \citep[Simulating EAGLE LEnses, ][]{Mukherjee2018, Mukherjee2021, Mukherjee2022} managed to constrain the stellar and AGN feedback processes that underlie the galaxy formation of massive lens galaxies, by simulating and modeling strong lenses from the EAGLE suite of cosmological simulations \citep[Evolution and Assembly of GaLaxies and their Environments, ][]{Crain2015, Schaye2015}. The programme obtained the projected dark matter (DM) fractions within both half the effective radius and the effective radius of simulated galaxies, showing good agreement with SLACS \citep[Sloan Lens ACS Survey, ][]{Bolton2006} lenses results.
In recent years, cosmological simulation suites featuring thousands of different combinations of cosmological and astrophysical parameters have emerged, such as \textsc{camels} \citep{Villaescusa-Navarro2021,Villaescusa-Navarro2023,Ni2023}. These suites are a powerful tool, in that they allow us to explore a wide range of parameter values, showing how cosmology and astrophysics influence the statistical properties of galaxies. In fact, it has been shown that one can infer information about cosmology and astrophysics even just from the physical properties of a single galaxy, by training a machine learning model on simulated galaxies \citep{Villaescusa-Navarro2022_one_galaxy,Rojas2023}, with an increased constraining potential when using the properties of multiple galaxies \citep{Chawak2024}. Other approaches that utilise \textsc{camels} to infer cosmological parameters are the use of graph neural networks for field-level likelihood-free inference of $\Omega_{\textrm{m}}$ with 12 per cent of precision \citep{deSanti2023} and the inference of both $\Omega_{\textrm{m}}$ and $\sigma_{8}$ with galaxy photometry alone \citep{Hahn2024}. In the context of cosmological simulations, the work by \cite{Wu2024} shows that it is also possible to infer both total and dark matter mass within the effective radius of real galaxies by training a machine learning algorithm on photometry, sizes, stellar mass and kinematic features of simulated galaxies, with very high accuracy and almost no bias.

In this context, galaxy scaling relations -- the relationships between different observable properties of galaxies -- have been proven to be a useful tool to probe the effects of different astrophysical feedback processes \citep{Tortora2019}. In \cite{Busillo2023}, hereafter referred to as Paper I, we developed a statistical method to determine, from a selection of \textsc{camels} cosmological simulations with different values of cosmological and astrophysical parameters, which simulation's set of galaxy scaling relations better fits some observed scaling relation for SPARC late-type galaxies \citep[LTGs, ][]{Lelli2016}.

In addition to gravity, cosmological simulations include various models for physical processes, such as Supernovae and AGN feedback. These processes drive the conversion of gas into stars, and their efficiency is known to vary with mass \citep{Dekel2004, Tortora2019, Hunt2020}. Furthermore, the distribution and the amount of dark matter and the expansion history of the Universe (summarised in the standard cosmological model) influences the distribution of the primordial dark matter seeds, the merger history of galaxies, and the mass assembly \citep{Taylor1992, Conselice2014}. 
A comprehensive comparison of observed scaling relations, involving stars and dark matter, across a broad range of masses, galaxy types and data samples with simulations is imperative. Therefore, in this Paper II the aim is to complement the analysis conducted in Paper I. We initially explore the scaling relations for massive early-type galaxies\footnote{For simplicity, we ignore the complexity of galaxy classification based on morphology, colour or star formation activity, each of which can bring to different galaxy selections. Therefore, throughout this paper, we will use "early-type galaxies" as a synonym for "passive galaxies", and "late-type galaxies" as a synonym for "star-forming galaxies".} (ETGs) as a function of astrophysical and cosmological parameters. Subsequently, we compare these relations with observations, including in the analysis the scaling relations for LTGs. Our aim is to describe the dark matter content of galaxies across different galaxy types and a mass range of approximately three orders of magnitude in stellar mass. 

The paper is structured as follows. In Section \ref{sec:Data}, we will present both the simulated and observational datasets used in our analysis. In Section \ref{sec:preliminary_discussion}, we discuss some of the properties of the ETGs that we take from \textsc{camels} for our analysis. We compare both the \textsc{camels} simulations and the original IllustrisTNG simulations with the various observational datasets in Section \ref{sec:results}, and state our conclusions in Section \ref{sec:conclusions}.

\section{Data}\label{sec:Data}
In this section, we will introduce both the simulated and the observational datasets used for the analysis. We begin by describing the simulated dataset used, which includes data from both \textsc{camels} and the `original' IllustrisTNG. We then describe the three main observational dataset used: SPIDER, \ATLAS\ and MaNGA DynPop. 
For each dataset, we list the main observational quantities used for the analysis, and outline any filtering process operated on the raw datasets.

\begin{table*}
    \centering
\caption{List of spatial, mass and numerical resolution parameters for the simulations with the best resolution, i.e. TNG300-1, TNG100-1 and TNG50-1, two more simulations with the same box side-length of TNG100-1, but worse resolution, i.e. TNG100-2 and TNG100-3, and \textsc{camels}.}
\label{tab:IllustrisTNG_resolution_table}
    
    \begin{tabular}{cccccccc} 
    \hline
    \hline
         Sim. Name&  $L_{\textrm{box}}/\mathrm{cMpc}$&  $N_{\textrm{gas}}$&  $N_{\textrm{DM}}$&  $m_\textrm{baryon}/M_{\odot}$&  $m_{\textrm{DM}}/M_{\odot}$&  $\epsilon_{\textrm{DM, stars}}^{z=0}/\mathrm{kpc}$&  $\epsilon_{\textrm{gas, min}}/\mathrm{ckpc}$\\
    \hline
         TNG300-1&  302.6&  $2500^{3}$&  $2500^{3}$&  $1.1\times 10^{7}$&  $5.9\times 10^{7}$&  1.48&  0.369\\
         TNG100-1&  106.5&  $1820^{3}$&  $1820^{3}$&  $1.4\times 10^{6}$&  $7.5\times 10^{6}$&  0.74&  0.185\\
 TNG100-2& 106.5& $910^{3}$& $910^{3}$& $1.1\times 10^{7}$& $6.0\times 10^{7}$& 1.48&0.369\\
 TNG100-3& 106.5& $455^{3}$& $455^{3}$& $8.9\times 10^{7}$& $4.8\times 10^{8}$& 2.95&0.738\\
         TNG50-1&  51.7&  $2160^{3}$&  $2160^{3}$&  $8.5\times 10^{4}$&  $4.5\times 10^{5}$&  0.29&  0.074\\
 \textsc{camels}& 37.3& $256^{3}$& $256^{3}$& $1.89\times 10^{7}$& $3.85(\Omega_{\textrm{m}}-\Omega_{\textrm{b}})\times 10^{8}$& 2.00&$-$\\
 \hline
    \end{tabular}
\tablefoot{The parameters are, from left to right: box side-length, initial number of gas cells and DM particles, baryon particle mass, DM particle mass, $z=0$ gravitational softening length of the DM particles and minimum comoving value of the adaptive gas gravitational softening length. Values are taken from \protect\cite{Nelson2019b}. The value of $\epsilon_{\textrm{gas, min}}$ for \textsc{camels} is unavailable, but it is close to the value for IllustrisTNG300-1.}
\end{table*}

\subsection{Simulated data}
In the following, we provide a detailed description of the simulated datasets employed in our analysis. We first focus on the IllustrisTNG suite of \textsc{camels}, detailing its main characteristics and the quantities that we chose to use in our analysis. Next, we describe the `original' simulation suites from the IllustrisTNG project, specifying what simulations have been considered for the analysis and the values of the cosmological parameters assumed in them.

\subsubsection{\textsc{camels}}\label{sec:Camels}
Similarly to Paper I, we make use of the simulated galaxy data coming from \textsc{camels}, a suite of cosmological simulations of an Universe volume equal to $25\,h^{-1}\;\textrm{Mpc}^{3}$ \citep{Villaescusa-Navarro2021, Villaescusa-Navarro2023, Ni2023}. The specific details of the simulations are thoroughly described in Paper I and are summarized in Table \ref{tab:IllustrisTNG_resolution_table}. For our purposes, we will only report the values of the cosmological parameters which are fixed for each simulation: $\Omega_{\textrm{b}} = 0.049$, $n_{\textrm{s}} = 0.9624$ and $h = 0.6711$. It's worth noting that these parameters, which are held constant in our analysis, exhibit degeneracies with other cosmological parameters, such as $\Omega_{\textrm{m}}$ and $\sigma_{8}$. The simulations used in this study do not allow for variations of these parameters.  Future works may utilise new \textsc{camels} simulations, which vary up to 28 parameters simultaneously, including $\Omega_{\textrm{b}}$, $n_{\textrm{s}}$ and $h$, both traditional \citep{Ni2023} and zoom-in \citep{Lee2024}. These new simulations, moreover, possess other supernovae feedback parameters than just $A_{\textrm{SN1}}$ (which, in IllustrisTNG, regulates wind energy per unit star formation rate) and $A_{\textrm{SN2}}$ (which instead regulates the wind velocity at ejection). They also provide other AGN feedback parameters than $A_{\textrm{AGN1}}$ and $A_{\textrm{AGN2}}$ (which regulate the energy per unit BH accretion rate and the burstiness of the kinetic-mode AGN feedback, respectively). We also emphasise that, while the \textsc{camels} simulation suites are not calibrated per se, the fiducial simulations use the same parameters as the respective original models, which instead have been calibrated. In particular, for IllustrisTNG, the observables on which the original simulation has been calibrated are described in Section \ref{sec:Original_IllustrisTNG}.

For this work, we primarily used simulated galaxy data coming from the 1061 `LH' and `1P' simulations of the IllustrisTNG suite \citep{Villaescusa-Navarro2021}. For the comparison with observations, we made use of the following quantities, obtained from \textsc{subfind} outputs, relative to the $z=0$ snapshot:

\begin{enumerate}
    \item Stellar half-mass radius, $R_{*,1/2}$;
    \item Total stellar mass, $M_{*}$;
    \item Total mass, $M_{\textrm{tot}}$;
    \item Stellar/DM/total mass within the half mass radius, $M_{*,1/2}$, $M_{\textrm{DM},1/2}$, and $M_{1/2}$, respectively;
    \item  DM fraction within the stellar half-mass radius, $f_{\textrm{DM}}(<R_{*,1/2}) \equiv  M_{\textrm{DM},1/2}/M_{1/2}$;
    \item Number of star particles within the stellar half-mass radius, $N_{*,1/2}$;
    \item Star formation rate, $\textrm{SFR}$.
\end{enumerate}
For the specific definition of these 3D quantities, see Paper I. 

Similarly to the previous work, we performed a filtering of the subhalos detected by \textsc{subfind}. We considered only subhalos that have $R_{*,1/2}>\epsilon_{\textrm{min}}$, $N_{*,1/2}>50$ and $f_{\textrm{DM}}(<R_{*,1/2}) > 0$, where $\epsilon_{\textrm{min}}=2\; \textrm{ckpc}$ is the gravitational softening length of the IllustrisTNG suite. Similarly to Paper I, we follow \cite{Bisigello2020} for the selection of ETGs via specific SFR ($\textrm{sSFR}:=\textrm{SFR}/M_{*}$), choosing for the ETGs only subhalos having $\log_{10}(\textrm{sSFR}/\textrm{yr}^{-1})\leq-10.5$, and the opposite for LTGs. Similarly to the results of Paper I, we have verified that varying the $\textrm{sSFR}$ threshold by $\pm 0.5\,\textrm{dex}$ does not affect our results. Notice that there will inevitably be some slight discrepancies between the observational samples of ETGs and LTGs and those derived from simulations, given that for the simulations we have performed a cut in specific star-formation rate for the selection, whereas the observational samples are selected based on different kinds of photometric and spectroscopic cuts. We checked that performing a photometric cut \citep{Tortora2010} instead of a sSFR cut produces a negligible variation in the median trends of the considered scaling relations.

\subsubsection{IllustrisTNG}\label{sec:Original_IllustrisTNG}
Apart from the \textsc{camels} simulated data, in order to analyse the impact of the simulation volume and resolution, we also made use of the `original' data from the IllustrisTNG project \citep{Pillepich2018, Nelson2018, Springel2018, Naiman2018, Marinacci2018, Pillepich2019, Nelson2019}. IllustrisTNG consists of 18 simulations in total, for three cubic volumes of the Universe.

For our work, we considered the three simulations with the highest resolution levels available for all three box sizes: TNG300-1, TNG100-1 and TNG50-1. We also used the low-resolution simulations of TNG100, TNG100-2 and TNG100-3, to check for resolution effects. Table \ref{tab:IllustrisTNG_resolution_table} sums up the spatial and mass resolution of the simulations used, along with the number of gas cells/DM particles contained within them.

The cosmological parameters associated with all three simulations are the following: $\Omega_{\textrm{m}} = 0.3089$, $\Omega_{\Lambda} = 0.6911$, $\Omega_{\textrm{b}} = 0.0486$, $\sigma_{8} = 0.8159$, $n_{\textrm{s}} = 0.9667$, $h = 0.6774$. These values show only minor deviations from the parameters of the fiducial \textsc{camels} IllustrisTNG simulation parameters (up to a 3 per cent difference for the cosmological parameter $\Omega_{\textrm{m}}$). The calibration of IllustrisTNG was performed by using the galaxy stellar mass function, the stellar-to-halo mass relation, the total gas mass content within the virial radius $r_{500}$ of massive groups, the stellar mass-stellar size and the black hole mass - galaxy mass relations, all at $z=0$. Additionally, the functional shape of the cosmic star formation rate density for $z\lesssim 10$ has been used \citep{Pillepich2018}.

For the analyses performed with these simulations, we considered the exact same quantities defined in Section \ref{sec:Camels} for the \textsc{camels} simulations, given that they share identical definitions with the IllustrisTNG suite. We filtered the catalogs in the same way as we have done for the subhalos in Section \ref{sec:Camels}, with the addition of $\texttt{SubhaloFlag} = 1$, where $\texttt{SubhaloFlag}$ is a flag given in the IllustrisTNG catalogs for the subhalos found by \textsc{subfind} which tells if a subhalo is of cosmological origin or not, i.e. it may have formed within an existing halo, is a disk fragment, etc. Typically, subhalos with $\texttt{SubhaloFlag} = 0$ are considered unsuitable for analysis, and thus we excluded them.

\subsection{Observational datasets}
In the following, we describe the observational datasets utilised to compare with the simulated data. We first introduce the SPIDER and \ATLAS datasets, which provide extensive data on early-type galaxies and their structural properties. We finally discuss the MaNGA DynPop dataset, comprising both early- and late-type galaxies, serving as a homogeneous sample for comparing trends of both simulated ETG and LTG galaxies. We detail the specific filtering criteria applied to distinguish between these two samples.
The fact that the three observational samples have different selection criteria provides insight into the impact of these selections on the final results, allowing us to explore the diversity among the observational samples. A description of some of the observational biases that are associated with the use of these observational datasets is reported in Appendix \ref{sec:observational_biases}.

Remarkably, except for the observational quantities derived by using direct distance measurements (e.g. SPARC velocity curves), the observational datasets are not cosmology-agnostic. Nevertheless, the impact of this dependency is negligible, given that the correction factor is at most of $0.05\,\textrm{dex}$ for the effective radii and total masses, and $0.10\,\textrm{dex}$ for the stellar masses. The effect of this shift on our scaling relations is negligible.

\subsubsection{SPIDER}
The first observational dataset that we considered is the Spheroids Panchromatic Investigation in Different Environmental Regions (SPIDER, \citealt{LaBarbera2010}) dataset, a sample of ETGs selected from the Sloan Digital Sky Survey Data Release 6 (SDSS-DR6). The sample is volume-limited, and covers a redshift range $z\in [0.05, 0.095]$, with available \textit{ugriz} photometry (also \textit{YJHK} photometry for a subsample of galaxies from UKIDSS-Large Area Survey-DR2) and optical spectroscopy. The galaxies have been selected such that $^{0.1}M_{\textrm{r}} < -20$, where $^{0.1}M_{\textrm{r}}$ is the $k$-corrected SDSS Petrosian magnitude in the $r$ band. The SPIDER sample is 95 per cent complete at stellar mass $M_{*} = 3\times 10^{10}\;M_{\odot}$, for a Chabrier IMF.

In this work, we considered the selection performed by \cite{Tortora2012}, in which only galaxies with high-quality structural parameters (Sérsic fit with $\chi^{2} < 2$ in all wavebands, uncertainty on $\log_{10}(R_{\textrm{eff}}/\textrm{kpc}) < 0.5\;\textrm{dex}$ from \textit{g} through \textit{K}) and available stellar mass estimates \citep{Swindle2011} have been selected. 

The galaxy properties which we consider are the following:
\begin{itemize}
    \item deprojected stellar half-light radius, $R_{*,1/2}$, derived from the (projected) effective radius $R_{\textrm{e}}$ (which is obtained from the Sérsic fit to the SDSS imaging in the $K$ band) by using the relation $R_{*,1/2}\approxeq 1.35\, R_{\textrm{e}}$ (\citealt{Wolf2010}, appendix B);
    \item stellar mass, $M_{*}$, obtained by fitting synthetic stellar population models from \cite{Bruzual2003}, using SDSS (Optical) + UKIDS (NIR) using the software \textsc{lephare} \citep{Ilbert2006}, assuming an extinction law \citep{Cardelli1989} and assuming a Chabrier IMF \citep{Chabrier2003};
    \item deprojected stellar mass within the half-light radius, $M_{*,1/2}$, obtained by halving $M_{*}$;
    \item dynamical mass within the stellar half-light radius, $M_{\textrm{dyn},1/2}$, obtained by modeling each galaxy using the Jeans equations, assuming a singular isothermal sphere (SIS) for the mass profile \citep{Tortora2012}.
\end{itemize}

From these quantities, we obtained the DM fraction within the stellar half-light radius, $f_{\textrm{DM}}(<R_{*,1/2})$, fixing it to zero whenever it becomes negative due to observational uncertainties. Similarly, we obtained the DM mass within the half-light radius by subtracting $M_{*,1/2}$ from $M_{\textrm{dyn},1/2}$, and fixing $M_{\textrm{DM},1/2} = M_{*,1/2}$ for those galaxies in which $f_{\textrm{DM}}(<R_{*,1/2}) \leq 0$.

The sample consists of 4260 ETGs, with more than 99 per cent of them residing in the red sequence, having $g-r\gtrsim 0.5$ within an aperture of $1\;R_{\textrm{eff}}$ and a median of $g-r = 0.88$.

\subsubsection{$\textrm{ATLAS}^{\textrm{3D}}$}
Our second sample consists of 258 ETGs from the ATLAS$^{3D}$ survey (\citealt{Cappellari2011,Cappellari+13_ATLAS3D_XV,
Cappellari+13_ATLAS3D_XX}). We have used this sample in our previous publications, where more details can be found \citep{Tortora+14_MOND, Tortora+14_DMslope}.  Our analysis is based on: 

\begin{itemize}
    \item $r$-band effective radius, $R_{\textrm{e}}$, used to derive the stellar half-light radius by using the relation $R_{*,1/2} \approxeq 1.35\,R_{\textrm{e}}$, similarly to the SPIDER sample;
    \item stellar mass, $M_{*}$, obtained by multiplying  the stellar mass-to-light ratio ($\Upsilon_*$), derived by fitting galaxy spectra with \cite{Vazdekis+12} single SSP MILES models, having a \cite{Salpeter55} IMF, and the $r$-band total luminosity $L_{r}$. Stellar masses are converted to a \cite{Chabrier01} IMF, considering the fact that the normalization of the Chabrier IMF is $\sim 0.25$ dex smaller than the Salpeter one. More details in \cite{Tortora+14_DMslope, Tortora+14_MOND}.
    \item stellar mass within $R_{*,1/2}$, $M_{*,1/2}$, obtained similarly to the SPIDER sample by halving the total stellar mass, $M_{*}$, based on a Chabrier IMF;
    \item dynamical mass within $R_{*,1/2}$, $M_{\textrm{dyn},1/2}$, derived via the same Jeans modeling applied to the SPIDER sample, by adopting a SIS profile and using the projected stellar velocity dispersion, $\sigma_{\textrm{e}}$\footnote{The Jeans dynamical modelling of SPIDER galaxies is based on velocity dispersions measured within SDSS fibers, which have a fixed radius which is obviously different from the effective radii of each individual galaxies, requiring some extrapolation to the effective radius in the mass calculation, while in the ATLAS$^{3D}$ no extrapolation is necessary.}, within $R_{\textrm{e}}$.
\end{itemize}

 The quantities $M_{\textrm{DM},1/2}$ and $f_\textrm{DM}(<R_{*,1/2})$ for the \ATLAS\ sample are then defined in the same way as for the SPIDER sample.
 
\subsubsection{MaNGA DynPop}
The third observational dataset comes from Mapping Nearby Galaxies at APO (MaNGA, \citealt{Bundy2015}), a survey which contains 3D spectroscopy of $\sim 10^{4}$ nearby galaxies. MaNGA provides two-dimensional maps for stellar velocity, stellar velocity dispersion, mean stellar age and star formation history for all the galaxies of the survey. Given that no preliminary selection on size, morphology or environment are applied on this catalog, MaNGA is a volume limited sample which is fully representative of the local universe galaxy population.

In this work, we made use of the DynPop catalog  \citep{Zhu2023, Lu2023}, which combines a stellar dynamics analysis performed using the Jeans Anisotropic Modeling (JAM) method \citep{Cappellari2008, Cappellari2020} with a stellar population synthesis method based on the Penalized Pixel-Fitting (pPXF, \citealt{Cappellari2004, Cappellari2017, Cappellari2023}) software. In the catalog, the JAM modeling part is formed by eight different set-ups. For our analysis, we made use of the spherically-aligned velocity ellipsoid ($\textrm{JAM}_{\textrm{sph}}$) plus generalized Navarro-Frenk-White (gNFW, \citealt{Wyithe2001}) halo density profile set-up. We have confirmed that this choice does not impact the results, as the difference in the values of total stellar mass, half-light radius, DM fraction and dark matter mass within the half-light radius across the eight set-ups is lower than $0.02\,\textrm{dex}$, well within the uncertainties.

We considered the following quantities to build the observational dataset:
\begin{itemize}
\item half-light radius (measured in $\textrm{kpc}$), the 3D radius of the sphere which encloses half of the total luminosity of the galaxy, converted from the listed quantity $\textrm{rhalf\_arcsec}$ (measured in $\textrm{arcsec}$) by using the respective angular-diameter distance (listed as $\textrm{DA}$);
\item total stellar mass, obtained from the total luminosity given by the MGE fitting (listed as $\textrm{Lum\_tot\_MGE}$ in the catalog) by multiplying it with the respective averaged intrinsic stellar mass-to-light ratio within the elliptical half-light isophote, $\textrm{ML\_int\_Re}$. A correction of $0.25\;\textrm{dex}$ was applied to the logarithm of the resultant stellar mass, to adjust for the universal Chabrier IMF assumed by \textsc{camels};
\item stellar mass within the half-light radius, obtained by halving the total stellar mass (by definition of half-light radius and how we linked the total stellar mass-to-the total luminosity);
\item total mass within the half-light radius, listed in the catalog as $\textrm{log\_Mt\_rhalf}$;
\item quality, Qual, which gives the visual quality of JAM models as -1, 0, 1, 2, 3 (from worst to best);
\item redshift of the galaxy, $z$;
\end{itemize}

Dark matter mass within the half-light radius and dark matter fraction within the half-light radius are then defined in the same way as for the SPIDER sample.

\begin{figure*}
    \centering
    \includegraphics[width=1\textwidth]{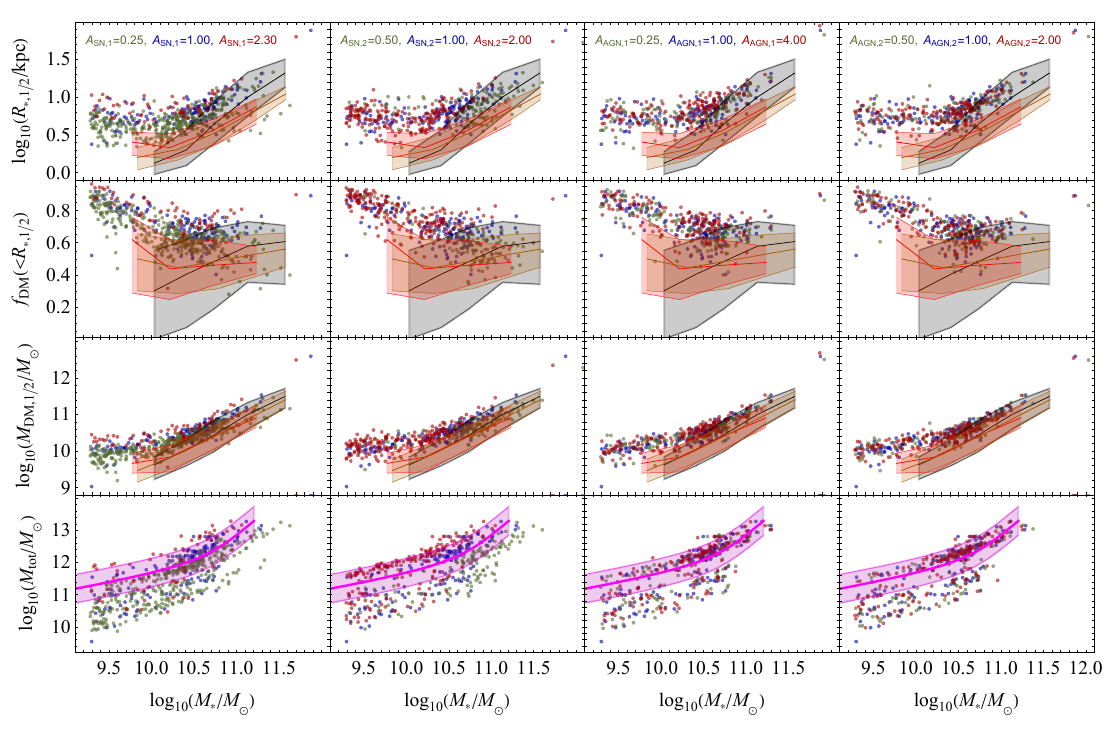}
    \caption{Comparison between SPIDER (grey region), \ATLAS\ (red region), MaNGA DynPop (orange region) and the theoretical $M_{\textrm{tot}}$-$M_{*}$ relation from \protect\cite{Moster2013} (pink region) with IllustrisTNG simulations having differing astrophysical feedback parameters. Each row in the plot corresponds to a different scaling relation, from top to bottom: stellar half-mass radius, $R_{*,1/2}$, DM fraction within $R_{*,1/2}$, $f_{\textrm{DM}}(<R_{*,1/2})$, DM mass within $R_{*,1/2}$, $M_{\textrm{DM},1/2}$, and total (virial) mass, $M_{\textrm{tot}}$, as a function of stellar mass, $M_{*}$. Each column shows the effect of varying one of the four astrophysical parameters. The continuous colored lines associated to the colored regions represent the 16th, 50th (median) and 84th percentile of the respective observational dataset distributions. The scatter of the \protect\cite{Moster2013} theoretical relation is taken to be the mean of the scatter of the $M_{\textrm{tot}}$-$M_{*}$ relation from \protect\cite{Posti2019}.}
    \label{fig:ETG_astrophysical_parameters_variation}
\end{figure*}

From the full sample of 10296 galaxies, we selected only the galaxies which respect the following properties:
\begin{itemize}
    \item $\textrm{Qual}\geq 1$, i.e. a good fit to either the velocity map, the velocity dispersion map, or both;
    \item $z\leq 0.1$, to remain consistent with the other observational datasets and with the fact that we considered a CAMELS snapshot at redshift $z = 0$.
\end{itemize}

To separate the ETGs from the LTGs, we made use of the MaNGA Morphology Deep Learning DR17 catalog \citep{Sanchez2022}. We classified as ETGs all the galaxies that exhibit the following properties:
\begin{itemize}
    \item $\textrm{T-Type} < 0$;
    \item $P_{\textrm{LTG}} < 0.5$;
    \item $\textrm{VC}=1$ or $\textrm{VC} = 2$;
    \item $\textrm{VF} = 0$.
\end{itemize}

Here, $P_{\textrm{LTG}}$ is a machine learning output indicating the probability that a galaxy in MaNGA is an LTG galaxy. Parameters $\textrm{VC}$ and $\textrm{VF}$ are associated with visual inspection, for a more robust classification: $\textrm{VC}$ indicates the visual class (where $\textrm{VC} = 1$ for ellipticals, $\textrm{VC} = 2$ for $\textrm{S0}$ and $\textrm{VC} = 3$ for spirals), and $\textrm{VF}$ denotes the visual flag (where $\textrm{VF} = 0$ signifies a reliable visual classification). The final sample thus consists of 1915 ETGs, which we used for our analysis.

We also considered the complementary LTG sample, taking all the galaxies with:
\begin{itemize}
    \item $\textrm{T-Type} \geq 0$;
    \item $P_{\textrm{LTG}} \geq 0.5$;
    \item $\textrm{VC} = 3$;
    \item $\textrm{VF} = 0$.
\end{itemize}

In this case, the final sample of LTGs consists of 2834 galaxies.

\begin{figure}
    \centering
    \includegraphics[width=1\linewidth]{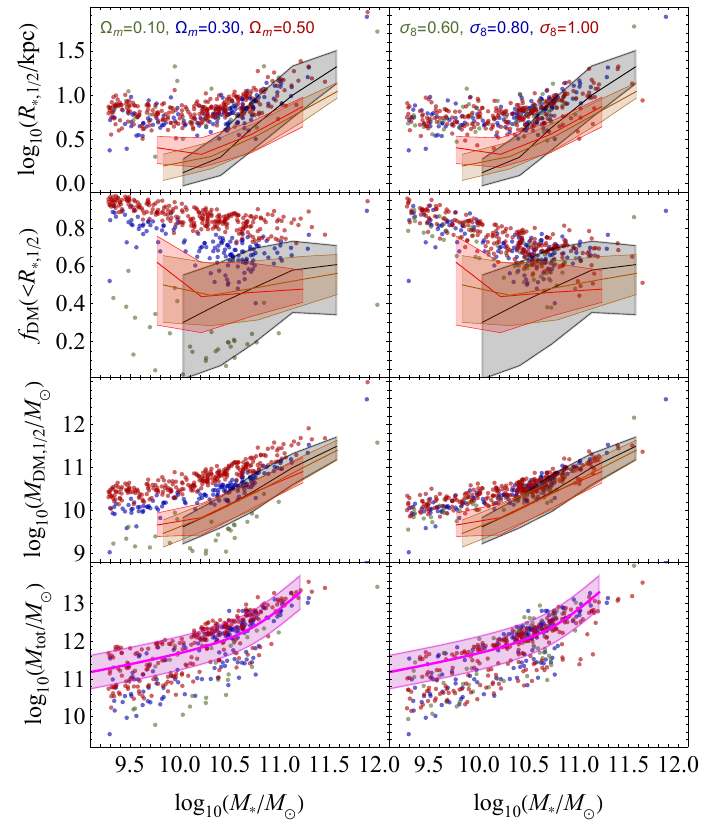}
    \caption{Same as Fig. \ref{fig:ETG_astrophysical_parameters_variation}, but considering the cosmological parameters $\Omega_{\textrm{m}}$ and $\sigma_{8}$.}
    \label{fig:ETG_cosmological_parameters_variation}
\end{figure}

\section{Properties of \textsc{camels} ETG sample}\label{sec:preliminary_discussion}
Before proceeding with the analysis, we first want to discuss the behavior of the ETGs in \textsc{camels}, to understand the key differences between simulated ETGs and the simulated LTGs analysed in Paper I. In Section \ref{sec:1P_comparison}, we examine how varying a single astrophysical or cosmological parameter affects the behavior of ETGs in regards to scaling relations. In Section \ref{sec:Number_galaxies_analysis} we analyse, for the first time, the differences between the number of LTGs and ETGs in each simulation as a function of the astrophysical and cosmological parameters.

\subsection{Comparison between observations and \textsc{camels} simulations}\label{sec:1P_comparison}

Fig. \ref{fig:ETG_astrophysical_parameters_variation} illustrates, for each column, the behavior of the four main scaling relations considered in this work (from top to bottom row: the 3D stellar half-mass radius, $R_{*,1/2}$, the internal DM fraction, $f_{\textrm{DM}}(<R_{*,1/2})$, the internal DM mass, $M_{\textrm{DM},1/2}$, and the total mass, $M_{\textrm{tot}}$, as a function of stellar mass, $M_{*}$). Each plot shows how the variation of individual astrophysical parameters impacts these relations, while keeping all other parameters fixed at their fiducial values.

We can see from Fig. \ref{fig:ETG_astrophysical_parameters_variation} that all simulated trends slightly overestimate the observations, except for the $M_{\textrm{tot}}$-$M_{*}$ relation, where the simulated points are largely compatible with the theoretical relation from \cite{Moster2013}. At fixed stellar mass, an increase in half-mass radius, dark matter fraction, dark matter mass within the half-mass radius and total mass is observed with increasing $A_{\textrm{SN1}}$, similarly to the LTG case studied in Paper I. Varying $A_{\textrm{SN2}}$ results in a slight increase in the number of high-mass galaxies as $A_{\textrm{SN2}}$ decreases. These galaxies consistently show systematically lower values of $R_{*,1/2}$, $f_{\textrm{DM}}(<R_{*,1/2})$ and $M_{\textrm{tot}}$ compared to galaxies in simulations with higher $A_{\textrm{SN2}}$ values. This contrasts with Paper I, where a decrease in $A_{\textrm{SN2}}$ led to higher values of $R_{*,1/2}$, $f_{\textrm{DM}}(<R_{*,1/2})$ and $M_{\textrm{DM},1/2}$. Nonetheless, the increase in the number of high-mass galaxies for lower $A_{\textrm{SN2}}$ values is less pronounced than observed in Paper I.

Interestingly, the effects of varying $A_{\textrm{AGN1}}$ and $A_{\textrm{AGN2}}$ are negligible even with high-mass early-type galaxies, contrary to what is expected in the literature from simulations, semi-analytical models and observations \citep{Dubois2016,DeLucia2017,DeLucia2024}. This, however, may be due to the fact that only these peculiar parameters for AGN feedback have negligible impact on the scaling relations for ETGs, and other parameters may influence the massive end of the scaling relations better than $A_{\textrm{AGN1}}$ and $A_{\textrm{AGN2}}$. Indeed, in the new \textsc{camels} simulations \citep{Ni2023}, other AGN-feedback parameters are avaliable, so in future works it may be possible to constrain these other parameters following an analogous procedure to the one used in this work.

Fig. \ref{fig:ETG_cosmological_parameters_variation} shows, instead, the effects of changing the cosmology on the four scaling relations. We can see that the effects are identical to the effects on LTGs studied in Paper I, where variations in $\Omega_{\textrm{m}}$ yield the greatest differences among simulations, while $\sigma_{8}$ has a negligible impact across simulations. This is to be expected, given that changing the values of cosmological parameters, e.g. $\Omega_{\textrm{m}}$, should affect the internal galaxy properties in a way that is independent of galaxy type as far as the direction of the shift is concerned, while specific differences due to the different assembly history between ETGs and LTGs produces different magnitudes of variations.

\subsection{Number of galaxies in the simulations}\label{sec:Number_galaxies_analysis}

\begin{figure*}
    \centering
    \includegraphics[width=1\linewidth]{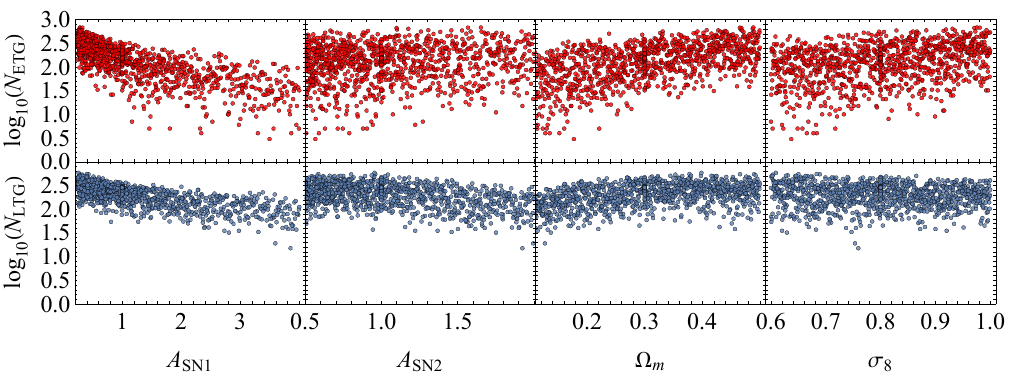}
    \caption{\textit{Top row:} Number of ETGs as a function of the cosmological/astrophysical parameters for each \textsc{camels} IllustrisTNG simulation. \textit{Bottom row}: Same as top row, but for LTGs. Each column is associated to a different parameter. For each plot, for a fixed value of one of the parameters, the trends with respect to the other parameters are the same as those shown in the other panels. Parameters $A_{\textrm{AGN1}}$ and $A_{\textrm{AGN2}}$ are not shown for convenience.}
    \label{fig:Number_galaxies_correlations}
\end{figure*}

In Paper I, our primary focus was to examine the dependence of scaling relations on astrophysical and cosmological parameters and compare them with observations. We did not specifically emphasise their distribution in terms of stellar mass. In this subsection, we investigate how the number of ETGs and LTGs, in a given simulation, changes as a function of the cosmological and astrophysical parameters. In particular, the number of ETGs in the simulations is systematically lower than the number of LTGs, mainly due to their higher mass and the relatively small cosmological volume of the \textsc{camels} simulations.

Fig. \ref{fig:Number_galaxies_correlations} shows the trends of the number of galaxies (ETGs and LTGs, respectively, on the top and bottom row) with respect to the most significant astrophysical parameters\footnote{We avoided considering the AGN feedback parameters for convenience, but we have checked that the trends of $N_{\textrm{LTG}}$ and $N_{\textrm{ETG}}$ with respect to $A_{\textrm{AGN1}}$ and $A_{\textrm{AGN2}}$ are constant.}. To assess monotonical correlations, we also conducted a hypothesis test for each parameter using Spearman's correlation coefficient as the test statistic, with a confidence level set at $99.7$ per cent. The Spearman correlation indexes, along with their corresponding p-values, are presented for both ETGs and LTGs in Table \ref{tab:Spearman_hypothesis_test}.

\begin{table*}
\renewcommand*{\arraystretch}{1.50}
\centering
\caption{Spearman test statistic, $\rho_{\textrm{s}}$, for the correlation shown in Fig. \ref{fig:Number_galaxies_correlations} between the number of  ETGs and LTGs, and each parameter.}
\label{tab:Spearman_hypothesis_test}
\begin{tabular}{ccccccc}
\hline
\hline
          & \multicolumn{3}{c}{ETGs}                                   & \multicolumn{3}{c}{LTGs}                                   \\ \hline
     Parameter              & $\rho_{\textrm{s}}$ & p-value               & significant? & $\rho_{\textrm{s}}$ & p-value               & significant? \\
\hline
$A_{\textrm{SN1}}$    & $-0.78$ & $1.01\times 10^{-213}$ & yes & $-0.72$ & $6.17\times 10^{-172}$ & yes \\
$A_{\textrm{SN2}}$ & $0.08$              & $6.47\times 10^{-3}$  & no           & $-0.36$             & $6.18\times 10^{-34}$ & yes          \\
$\Omega_{\textrm{m}}$ & $0.53$  & $7.15\times 10^{-79}$  & yes & $0.41$  & $1.05\times 10^{-43}$  & yes \\
$\sigma_{8}$       & $0.31$              & $2.25\times 10^{-24}$ & yes          & $-0.08$             & $8.19\times 10^{-3}$  & no     \\ \hline
\end{tabular}
\tablefoot{The associated p-value and the result of the hypothesis test are also reported for both types of galaxies.}
\end{table*}

From Fig. \ref{fig:Number_galaxies_correlations}, we observe that ETGs exhibit more scattered trends with respect to the corresponding LTG trends. For LTGs, both trends with $A_{\textrm{SN1}}$ and $A_{\textrm{SN2}}$ are decreasing, with $A_{\textrm{SN1}}$ showing a notably steeper decline. This is expected, given that a higher supernovae feedback corresponds to a lower number of late-type galaxies because of quenching. On the other hand, for ETGs, we find that the trend with $A_{\textrm{SN1}}$ is analogous to that of LTGs, while there is no trend with $A_{\textrm{SN2}}$, indicating that an increased energy per unit SFR has much more effect on quenching ETGs at $z=0$ than an increased wind velocity.

The trend of the number of both LTGs and ETGs with $\Omega_{\textrm{m}}$ is increasing, indicating that in a Universe with a larger DM content, more DM halos are formed. The trend with $\sigma_{8}$ is instead flat for LTGs, while for ETGs the trend is slightly increasing, possibly due to the fact that high-$\sigma_{8}$ cosmologies lead to an earlier formation of high-mass clusters, and as a consequence, a higher number of mergers inside the clusters.

\section{Comparison between simulations and observational datasets}\label{sec:results}

In this section, we first introduce an updated methodology for ranking simulations based on their closeness with observational datasets. Subsequently, we analyse the behavior of the fiducial and the IllustrisTNG best-fit simulation from Paper I with respect to the observational trends from SPARC,  \ATLAS\ and MaNGA DynPop. We proceed by identifying the best-fit simulations with respect to each of the three observational datasets, by using the procedure detailed in Section \ref{sec:ranking_method}. Lastly, we determine the best-fit simulations by considering both LTGs and ETGs in the observational samples, concluding with a comparison between the observed datasets and the original IllustrisTNG simulations.

For this analysis, we consider three different scaling relations: $R_{*,1/2}$-$M_{*}$, $f_{\textrm{DM}}(<R_{*,1/2})$-$M_{*}$ and $M_{\textrm{DM},1/2}$-$M_{*}$.

\subsection{Ranking of simulations}\label{sec:ranking_method}

Following Paper I, for each of the three observational datasets (SPIDER, \ATLAS, MaNGA DynPop) we rank how well each simulation fits the data by considering the cumulative reduced chi-squared, which is given by the sum of the reduced chi-squared for each of the three scaling relations considered in the analysis. In paper I, we defined the reduced chi-squared for a single scaling relation as:

\begin{equation}
\tilde{\chi}^{2}_{\textrm{rel}}=\frac{1}{N_{\textrm{sim}}-1}\sum_{i=1}^{N_{\textrm{sim}}}\frac{[y_{\textrm{sim, }i}-f_{\textrm{rel}}(x_{\textrm{sim, }i})]^{2}}{\sigma_{\textrm{rel},i}^2},
\end{equation}
\\
where $(x_{\textrm{sim},i},y_{\textrm{sim},i})$ are the $N_{\textrm{sim}}$ points from the simulation in the considered scaling relation parameter space, $f_{\textrm{rel}}$ is the observed scaling relation median trend's linear interpolation function, and $\sigma_{\textrm{rel},i}$ is given by the mean between $\sigma_{-}$ and $\sigma_{+}$, which are the differences, in absolute value, between the linear interpolated functions of the 16th and the 84th percentile trends associated to the observed scaling relation, respectively, and the interpolated median trend, each evaluated at $x_{\textrm{sim},i}$.
In this work, however, we use a slightly different definition for the reduced chi-squared:

\begin{equation}
\tilde{\chi}^{2}_{\textrm{rel}}=\frac{1}{N_{\textrm{sim}}-1}\,\sum_{i=1}^{N_{\textrm{sim}}}\frac{[y_{\textrm{sim, }i}-\mathcal{N}(f_{\textrm{rel}}(x_{\textrm{sim, }i}),\sigma_{\textrm{rel},i})]^{2}}{\sigma_{\textrm{rel},i}^2}. \label{eq:chi_squared_new_definition}
\end{equation}
The main difference between the new definition and the one used in Paper I is at the numerator, where this time we evaluate the difference between the $y$ component of the simulated point, $y_{\textrm{sim},i}$, and a randomly extracted point from a Gaussian distribution, $\mathcal{N}$, centered on the median of the observed scaling relation, with standard deviation equal to $\sigma_{\textrm{rel},i}$. This is done in order to obtain more realistic uncertainties.

In comparison to the previous paper, we also ensure that the simulations have realistic distributions in terms of the galaxy stellar mass. Specifically, we avoid situations where all the galaxies are clustered in a very tight range of stellar mass or where the distribution in terms of stellar mass is strongly discontinuous. To achieve this, we bin the quantities associated to the scaling relations, for example the stellar half-mass radius, with respect to values of stellar mass. We then filter out simulations that have less than one galaxy in each bin of stellar mass, using the same bin edges as those employed for the analogous binning of the observational trends. To lessen the constraint imposed by this filtering procedure, we exclude the very first and last bins with respect to stellar mass in this procedure.

We then evaluate the values of $\tilde{\chi}^{2}$ for all remaining \textsc{camels} simulations from the filtering procedure, and rank the simulations according to its value. We then consider the lowest $\tilde{\chi}^{2}$-value simulation as the best-fit simulation for a given observational dataset.

The uncertainty on the cosmological and astrophysical parameters is determined by performing 100 bootstrap resamplings of both the observational datasets and simulated data points, with each resampling being a copy of the original dataset having some of the galaxies substituted by copies of other galaxies in the same dataset. This is accomplished by using the \textsc{mathematica} resource function \textsc{BootstrapStatistics} (more informations are available in Paper I). Given a certain observational dataset, for each resampling, the simulations are ranked based on the value of $\tilde{\chi}^{2}$. By picking the best-fit simulation for each resampling, we obtain a sample of 100 best-fit simulations. We evaluate the empirical cumulative distribution functions (CDFs) for each cosmological and astrophysical parameter of this sample, via the formula:

\begin{equation}
\hat{F}_{n}(x) = \frac{\textrm{n° of elements in sample} < x}{n},
\end{equation}
\\
with $n = 100$ being the dimension of the sample. We then smooth the empirical CDFs with a Gaussian kernel having standard deviation $\sigma = 1$, in order to avoid issues deriving by the discrete nature of the empirical CDF. We then obtain the parameter estimates from the smoothed CDFs, by considering the associated 16th, 50th (median) and 84th percentiles. The results are listed in Table \ref{tab:bootstrap_results}. The constraints considering the 16th, 50th and 84th percentile taken directly from the empirical CDFs instead are detailed in Appendix \ref{tab:bootstrap_results_1_sigma}.

\begin{table*}
\centering
\renewcommand*{\arraystretch}{1.50}
\caption{Constraints on cosmological and astrophysical parameters, obtained by bootstrapping both observational and simulated datasets, and taking for each resampling the best-fit simulation.}
\label{tab:bootstrap_results}
    \begin{tabular}{ccccccccc}
    \hline
    \hline
         Obs. Trend&  $\Omega_{\textrm{m}}$&  $\sigma_{8}$&  $S_{8}$&  $A_{\textrm{SN1}}$&  $A_{\textrm{SN2}}$&  $A_{\textrm{AGN1}}$& $A_{\textrm{AGN2}}$ &$\tilde{\chi}^{2}$\\
    \hline
         SPIDER&  $0.25_{-0.05}^{+0.04}$&  $0.77_{-0.12}^{+0.13}$&  $0.67_{-0.10}^{+0.15}$&  $1.74_{-0.67}^{+0.78}$&  $0.64_{-0.11}^{+0.17}$&  $0.94_{-0.69}^{+1.84}$& $1.10_{-0.49}^{+0.40}$ &$3.52_{-0.27}^{+0.27}$\\
         \ATLAS&  $0.21_{-0.03}^{+0.02}$&  $0.87_{-0.18}^{+0.08}$&  $0.71_{-0.14}^{+0.07}$&  $0.31_{-0.05}^{+0.14}$&  $1.58_{-0.64}^{+0.40}$&  $1.05_{-0.64}^{+1.00}$& $1.02_{-0.45}^{+0.38}$ &$4.24_{-0.26}^{+0.26}$\\
         MaNGA DynPop&  $0.16_{-0.01}^{+0.03}$&  $0.95_{-0.14}^{+0.03}$&  $0.70_{-0.06}^{+0.02}$&  $0.30_{-0.03}^{+0.01}$&  $1.61_{-0.09}^{+0.16}$&  $2.57_{-1.74}^{+0.54}$& $1.53_{-0.91}^{+0.35}$ &$5.38_{-0.29}^{+0.34}$\\
    \hline
    \end{tabular}
\tablefoot{The values reported are the 16th, 50th (median) and 84th percentiles, taken from the respective empirical CDFs smoothed with a Gaussian kernel having standard deviation $\sigma = 1$. The last column shows the cumulative reduced chi-squared obtained for each simulation.}
\end{table*}

\subsection{\textsc{camels} fiducial simulation comparison with observations}\label{sec:Camels_fiducial_comparison}
As a preliminary check, we verify if the fiducial IllustrisTNG simulation from \textsc{camels} reproduces accurately the observational trends from SPIDER, \ATLAS\ and MaNGA DynPop for the ETG sample. The procedure to obtain the observational trends is the same as in Paper I.

\begin{figure}
    \centering
    \includegraphics[width=\linewidth]{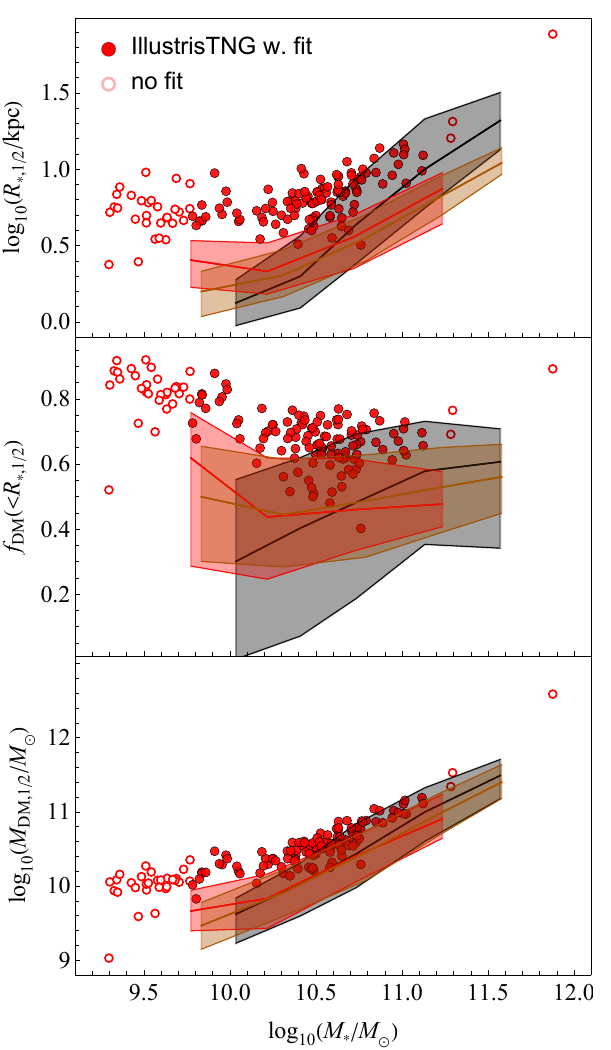}
    \caption{From top to bottom: stellar half-mass radius, $R_{*,1/2}$, DM fraction within $R_{*,1/2}$, $f_{\textrm{DM}}$, and DM mass within $R_{*,1/2}$, $M_{\textrm{DM},1/2}$, as a function of stellar mass, $M_{*}$, for the fiducial IllustrisTNG simulation `$\textrm{1P\_1\_0}$', compared with the corresponding SPIDER (black curves), \ATLAS\ (dark red curves) and MaNGA DynPop (orange curves) trends. The shaded areas represent the scatter of the observed relations, given by the difference between the 16th and the 84th percentiles with the median. As an example, the open circles are galaxies not used for the evaluation of the cumulative reduced chi-squared associated to the \ATLAS\ observational trend in Section \protect\ref{sec:Camels_fiducial_comparison}.}
    \label{fig:fiducial_ETG_trend}
\end{figure}

Fig. \ref{fig:fiducial_ETG_trend} shows the comparison between the simulated ETGs from the fiducial IllustrisTNG \textsc{camels} simulation, `$\textrm{1P\_1\_0}$', and the observational trends.

Quantitatively, the cumulative reduced chi-squared, $\tilde{\chi}^{2}$, is $\tilde{\chi}^{2} = 4.67$ with respect to SPIDER, $\tilde{\chi}^{2} = 7.65$ with respect to \ATLAS\ and $\tilde{\chi}^{2} = 11.14$ with respect to MaNGA DynPop. The associated ranks are equal to 316, 354 and 353, respectively, where a rank of 1 is associated to the best-fit simulation. This means that the fiducial simulation is far from being the better fit for all of the three observational trends.

Visually, one can see that there is a systematic overestimate of the considered quantities, at fixed stellar mass, with respect to the observational values. This finding aligns with the overestimation observed in Paper I for the fiducial IllustrisTNG simulation, concerning LTGs. However, for ETGs, the discrepancy is even more pronounced, with the simulated trends being compatible within $1\sigma$ only with the SPIDER observations, and only at high mass ($\log_{10}(M_{*}/M_{\odot}) \gtrsim 10.5$).

\subsection{Consistency of ETG results with IllustrisTNG LTG best-fit simulation}\label{sec:CASCO1_bestfit_comparison}

\begin{figure}
    \centering
    \includegraphics[width=\linewidth]{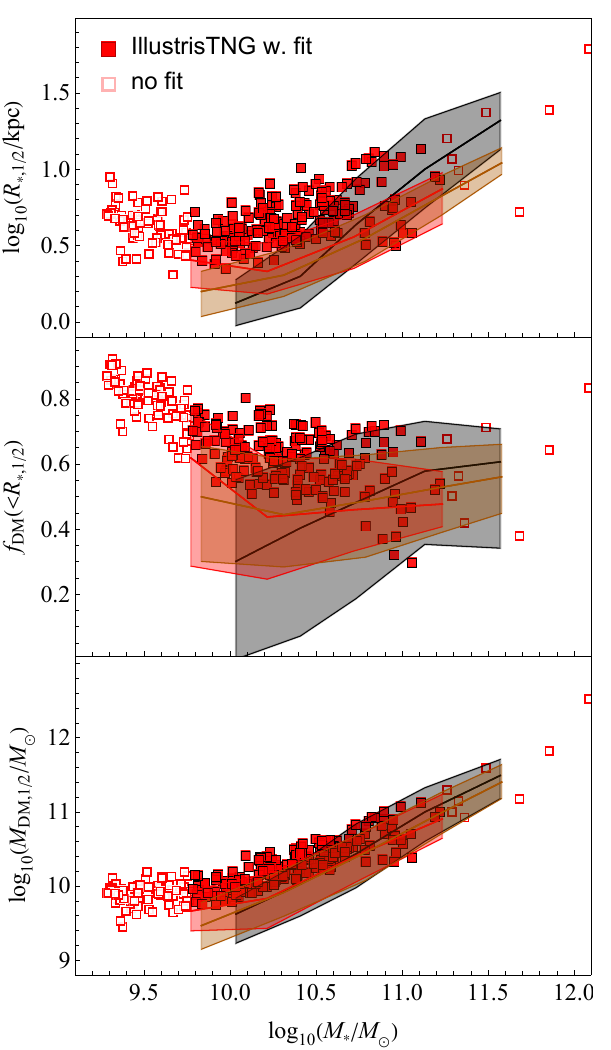}
    \caption{Same as Fig. \ref{fig:fiducial_ETG_trend}, but for the best-fit simulation `LH\_698', found in Paper I to be the best-fit simulation between the simulated LTG data and the SPARC trends.}
    \label{fig:CASCO1_ETG_trend}
\end{figure}

In Paper I, we found that the \textsc{camels} simulation `LH\_698', with parameters $\Omega_{\textrm{m}} = 0.27$, $\sigma_{8} = 0.83$, $S_{8} = 0.78$, $A_{\textrm{SN1}} = 0.48$, $A_{\textrm{SN2}} = 1.24$, $A_{\textrm{AGN1}} = 2.53$ and $A_{\textrm{AGN2}} = 1.79$, is the best-fit simulation reproducing the observed trends for LTGs from SPARC \citep{Lelli2016}. In Fig. \ref{fig:CASCO1_ETG_trend} we show the comparison between the simulated ETGs from this simulation and the observational trends from SPIDER, \ATLAS\ and MaNGA DynPop.

From Fig. \ref{fig:CASCO1_ETG_trend}, we can see that the simulation `LH\_698' is not reproducing the observed correlations for the ETG sample: indeed, for this simulation the cumulative reduced chi-squared is $\tilde{\chi}^{2} = 4.78$ with respect to SPIDER, $\tilde{\chi}^{2} = 4.48$ with respect to \ATLAS\ and $\tilde{\chi}^{2} = 7.35$ with respect to MaNGA DynPop, with an ordered rank of 324, 150 and 157, respectively.

While in the case of SPIDER the ranking is slightly higher than the ranking of the fiducial simulation, in the case of \ATLAS\ and MaNGA DynPop the ranking is much lower, showing a slightly better compatibility of this simulation with the observational datasets. The comparison between the values of the two simulations is reported in Table \ref{tab:fiducial_CASCO1_comparison}.

\begin{table}
\renewcommand*{\arraystretch}{1.50}
    \centering
\caption{Comparison between the cumulative reduced chi-squared for the fiducial IllustrisTNG `1P\_1\_0' simulation and the IllustrisTNG best-fit simulation `LH\_698' from Paper I.}
\label{tab:fiducial_CASCO1_comparison}
    
    \begin{tabular}{cccc}
    \hline
    \hline
         Sim. Name&  $\tilde{\chi}^{2}_{\textrm{SPIDER}}$& $\tilde{\chi}^{2}_{\textrm{\ATLAS}}$ &$\tilde{\chi}^{2}_{\textrm{MaNGA}}$\\
    \hline
         1P\_1\_0&  4.67&  7.65&11.14\\
        LH\_698& 4.78&4.48 &7.35\\
    \hline
    \end{tabular}

\end{table}

Visually, we can see from the figure that the simulated ETG sample systematically overestimates all three observational trends, except at high mass ($\log_{10}(M_{*}/M_{\odot}) \gtrsim 10.7$). Even at high mass, however, the improved alignment between simulated ETGs from the best-fit simulation of Paper I and observational trends is primarily attributed to the greater dispersion exhibited by the simulated data points.

\subsection{Best fit to the observations}
In Sections \ref{sec:Camels_fiducial_comparison} and \ref{sec:CASCO1_bestfit_comparison}, we have seen that neither the fiducial nor the best-fit simulation of Paper I accurately reproduce the observed trends from SPARC, \ATLAS\ and MaNGA DynPop. We thus proceed with the method detailed in Section \ref{sec:ranking_method}, to find the best-fit simulation for the various observational trends.

\begin{figure*}
    \centering
    \includegraphics[width=1\textwidth]{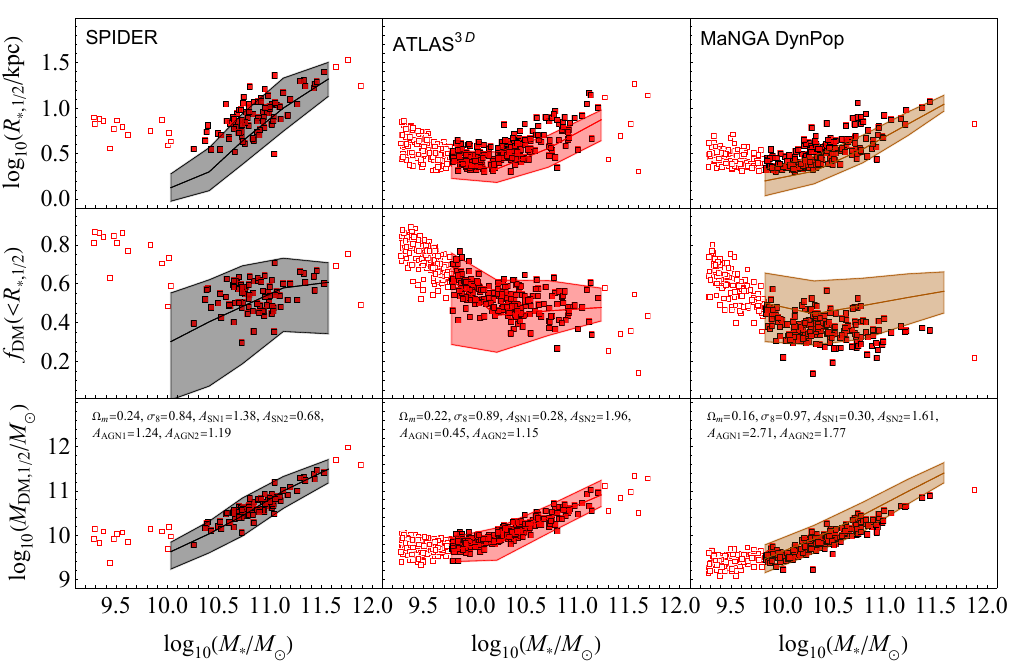}
    \caption{Each column of this figure is the same as Fig. \ref{fig:fiducial_ETG_trend}, but for the best-fit simulation associated to SPIDER (left column), \ATLAS\ (center column) and MaNGA DynPop (right column) observational trends. Red squares are the simulated galaxies. Filled red squares are for galaxies used for the evaluation of the cumulative reduced chi-squared with respect to the associated observational trend.}
    \label{fig:best_fit_trends}
\end{figure*}

\subsubsection{SPIDER}\label{sec:SPIDER_best_fit}
For the SPIDER trends, we find that the best-fit simulation is the simulation `LH\_523', having the following cosmological and astrophysical parameters: $\Omega_{\textrm{m}} = 0.24$, $\sigma_{8} = 0.84$, $S_{8} = 0.74$, $A_{\textrm{SN1}} = 1.38$, $A_{\textrm{SN2}} = 0.68$, $A_{\textrm{AGN1}} = 1.24$ and $A_{\textrm{AGN2}} = 1.19$, where the value of $S_{8}$ has been determined from the definition, $S_{8} := \sigma_{8}\,\sqrt{\Omega_{\textrm{m}}/0.3}$. The cumulative reduced chi-squared associated to this simulation is $\tilde{\chi}^{2} = 1.16$.

The first column of Fig. \ref{fig:best_fit_trends} illustrates the comparison between the ETGs of this simulation and the observed SPIDER trends. The figures reveals a slight overestimation by the simulated galaxies of the size-mass relation. The simulation also shows a lack of galaxies in the low-mass tail of the observational trends. This is especially evident in the size-mass relation.

Through bootstrap resampling, we derived the constraints for the cosmological and astrophysical parameters reported in the first row of Table \ref{tab:bootstrap_results}. Results show a slightly lower value of $\Omega_{\textrm{m}}$, compatible with \cite{DESandKiDS2023} and Paper I results, but not with \cite{Planck2018} within $1\sigma$. The values of $\sigma_{8}$ and $S_{8}$ are instead compatible with both results, along with Paper I. Regarding the astrophysical parameter $A_{\textrm{SN1}}$, SPIDER constraints suggest a higher estimate compared to the fiducial unit value from \textsc{camels}, incompatible with the results of Paper I that showed a value of $A_{\textrm{SN1}}$ equal to $0.48_{-0.16}^{+0.25}$ by comparing simulations with the SPARC LTG sample, which is instead lower than the fiducial unit value. Correspondingly, we see a lower value of $A_{\textrm{SN2}}$ than the fiducial value, also incompatible with Paper I results that show a value of $A_{\textrm{SN2}}$ equal to $1.21_{-0.34}^{+0.03}$. Constraints on $A_{\textrm{AGN1}}$ and $A_{\textrm{AGN2}}$ are not particularly significant, with $A_{\textrm{AGN1}}$ unconstrained and $A_{\textrm{AGN2}}$ compatible within the uncertainties with the fiducial value.
\begin{figure*}
    \centering
    \includegraphics[width=1\textwidth]{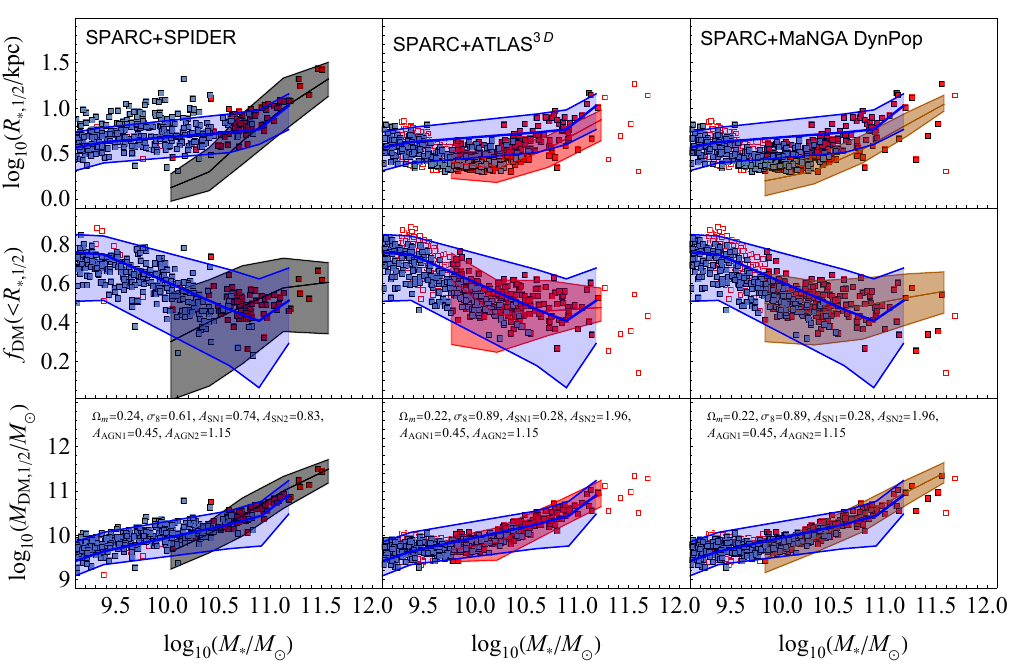}
    \caption{Same as Fig. \ref{fig:best_fit_trends}, but considering both simulated LTGs (blue points) and simulated ETGs (red points). The blue region is the SPARC observational trend from Paper I.}
    \label{fig:LTG_plus_ETG_best_fit_trends}
\end{figure*}
    
\subsubsection{$\textrm{ATLAS}^{\textrm{3D}}$}
For the trends in the \ATLAS\ sample, we find that the best-fit simulation is the simulation `LH\_797', having the following cosmological and astrophysical parameters: $\Omega_{\textrm{m}} = 0.22$, $\sigma_{8} = 0.89$, $S_{8} = 0.77$, $A_{\textrm{SN1}} = 0.28$, $A_{\textrm{SN2}} = 1.96$, $A_{\textrm{AGN1}} = 0.45$ and $A_{\textrm{AGN2}} = 1.15$, with a cumulative reduced chi-squared value of $\tilde{\chi}^{2} = 1.36$.

The central column of Fig. \ref{fig:best_fit_trends} shows the comparison between the ETGs of this simulation and the observed \ATLAS\ trends. From the figure, we can see that the simulated galaxies closely match the observed \ATLAS\ trends, albeit with a slight overestimate of the three quantities at fixed stellar mass, which is accentuated for the size-mass relation at the high-mass end.

Cosmological constraints from \ATLAS, reported in the second row of Table \ref{tab:bootstrap_results}, are compatible with constraints from SPIDER and Paper I within the uncertainties considered. The parameter $\Omega_{\textrm{m}}$, constrained from \ATLAS, is not compatible with both \cite{Planck2018} and \cite{DESandKiDS2023} results within $1\sigma$, but is closer to the latter. The constraint of $\sigma_{8}$ instead is compatible with both results, while the constraint of $S_{8}$ is only compatible with \cite{DESandKiDS2023} results. The values of $A_{\textrm{SN1}}$ and $A_{\textrm{SN2}}$ obtained show an inversion compared to the ones from SPIDER: $A_{\textrm{SN1}}$ is significantly lower, while $A_{\textrm{SN2}}$ is higher than the respective fiducial values, although the latter remains compatible with the fiducial value within the uncertainties. Both supernovae feedback parameters are compatible with the results found in Paper I.

\subsubsection{MaNGA DynPop}
For the MaNGA DynPop trends, we find that the best-fit simulation is the simulation `LH\_586', having the following cosmological and astrophysical parameters: $\Omega_{\textrm{m}} = 0.16$, $\sigma_{8} = 0.97$, $S_{8} = 0.70$, $A_{\textrm{SN1}} = 0.30$, $A_{\textrm{SN2}} = 1.61$, $A_{\textrm{AGN1}} = 2.71$ and $A_{\textrm{AGN2}} = 1.77$, with a cumulative reduced chi-squared value of $\tilde{\chi}^{2} = 2.53$.

The right column of Fig. \ref{fig:best_fit_trends} shows the comparison between the ETGs of this simulation and the observed MaNGA DynPop trends. From the figure, we can see that there is a low agreement between the best-fit simulation and the observed trends. In particular, there is a systematic overestimation of the size in the size-mass relation, and a systematic underestimate of dark matter mass in the $M_{\textrm{DM},1/2}$-$M_{*}$ relation, which in turn leads to a systematic underestimate of DM fraction in the $f_{\textrm{DM}}(<R_{*,1/2})$-$M_{*}$ relation.

The constraints for MaNGA DynPop are reported in the third row of Table \ref{tab:bootstrap_results}. In this case, the value of $\Omega_{\textrm{m}}$ is much lower than the ones obtained from the fitting procedure with SPIDER and \ATLAS\ observations, other than the results from Paper I, while the value of $\sigma_{8}$ is instead compatible with all of them within the uncertainties, albeit with a very high median value. This produces a very low value of $S_{8}$, not compatible with \cite{Planck2018} nor \cite{DESandKiDS2023} results. The value of $S_{8}$ is however compatible with the values inferred from SPIDER and \ATLAS\ observations, and from SPARC in Paper I, within the uncertainties. Supernovae feedback parameters are in agreement with the constraints found for \ATLAS. AGN feedback parameters are unconstrained.

\subsection{A best-fit solution for ETGs and LTGs}
\begin{table*}
\centering
\renewcommand*{\arraystretch}{1.50}
\caption{Constraints on cosmological and astrophysical parameters, obtained by bootstrapping the observational (including SPARC and MaNGA DynPop LTGs) and simulated datasets, and taking for each resampling the best-fit simulation.}
\label{tab:bootstrap_results_LTGs_plus_ETGs}
    \begin{tabular}{ccccccccc}
    \hline
    \hline
         Obs. Trend&  $\Omega_{\textrm{m}}$&  $\sigma_{8}$&  $S_{8}$&  $A_{\textrm{SN1}}$&  $A_{\textrm{SN2}}$&  $A_{\textrm{AGN1}}$& $A_{\textrm{AGN2}}$ &$\tilde{\chi}^{2}$\\
    \hline
         SPARC + SPIDER&  $0.24_{-0.03}^{+0.01}$&  $0.74_{-0.13}^{+0.11}$&  $0.66_{-0.12}^{+0.08}$&  $0.65_{-0.31}^{+0.17}$&  $0.78_{-0.11}^{+0.14}$&  $0.96_{-0.61}^{+1.19}$& $1.07_{-0.40}^{+0.31}$ &$6.33_{-0.41}^{+0.48}$\\
         SPARC + \ATLAS&  $0.21_{-0.01}^{+0.02}$&  $0.94_{-0.07}^{+0.04}$&  $0.78_{-0.04}^{+0.02}$&  $0.31_{-0.03}^{+0.09}$&  $1.23_{-0.30}^{+0.75}$&  $1.24_{-0.77}^{+0.29}$& $0.75_{-0.17}^{+0.48}$ &$6.45_{-0.41}^{+0.68}$\\
         SPARC + MaNGA DynPop&  $0.22_{-0.01}^{+0.01}$&  $0.89_{-0.07}^{+0.07}$&  $0.76_{-0.06}^{+0.03}$&  $0.31_{-0.04}^{+0.07}$&  $1.58_{-0.55}^{+0.44}$&  $1.02_{-0.70}^{+0.94}$& $1.03_{-0.43}^{+0.38}$ &$8.23_{-0.65}^{+0.71}$\\
 MaNGA DynPop ETGs + LTGs& $0.25_{-0.02}^{+0.05}$& $0.89_{-0.19}^{+0.07}$& $0.82_{-0.12}^{+0.04}$& $1.83_{-1.19}^{+0.74}$& $0.58_{-0.05}^{+0.02}$& $0.48_{-0.15}^{+1.09}$&$0.95_{-0.20}^{+0.21}$ &$14.37_{-0.85}^{+0.51}$\\
    \hline
    \end{tabular}
\tablefoot{The values reported are the 16th, 50th (median) and 84th percentiles, taken from the respective empirical CDFs smoothed with a Gaussian kernel having standard deviation $\sigma = 1$. The last column shows the cumulative reduced chi-squared obtained for each simulation.}
\end{table*}

In Section \ref{sec:Camels_fiducial_comparison}, we identified a systematic upwards shift of all three scaling relations in the fiducial \textsc{camels} simulation, with respect to the observational trends. 

In Section \ref{sec:CASCO1_bestfit_comparison}, we also identified that the simulation identified as the best-fit for SPARC in Paper I ranks low when comparing the corresponding ETG galaxies with observations. The main cause of this could be the fact that we did not constrain both LTGs and ETGs simultaneously: consequently, the procedure primarily aimed at finding the LTG population that best matched the observations, without ensuring that the simulation also accurately represented the behavior of ETGs.

We have thus repeated the chi-squared procedure by constraining both LTGs against SPARC and the ETGs against the three observational ETG trends (SPIDER, \ATLAS\ and MaNGA DynPop), then obtaining the cumulative reduced chi-squared for both ETGs and LTGs simultaneously, $\tilde{\chi}^{2}_{\textrm{LTGs+ETGs}}$, by evaluating the sum of the cumulative reduced chi-squared for each of the two galaxy types and ranking the simulations with respect to this sum.

We find that, for SPIDER, the best-fit simulation is the `LH\_325' simulation, while for both \ATLAS\ and MaNGA DynPop the best-fit simulation is the `LH\_797' simulation. The former has the following cosmological and astrophysical parameters: $\Omega_{\textrm{m}} = 0.24$, $\sigma_{8} = 0.61$, $S_{8} = 0.55$, $A_{\textrm{SN1}} = 0.74$, $A_{\textrm{SN2}} = 0.83$, $A_{\textrm{AGN1}} = 0.38$ and $A_{\textrm{AGN2}} = 1.02$, with a cumulative reduced chi-squared associated to SPIDER of $\tilde{\chi}^{2}_{\textrm{LTGs+ETGs}} = 3.43$. The latter, which is also the best-fit simulation with respect to \ATLAS\ alone, has instead the following parameters: $\Omega_{\textrm{m}} = 0.22$, $\sigma_{8} = 0.89$, $S_{8} = 0.77$, $A_{\textrm{SN1}} = 0.28$, $A_{\textrm{SN2}} = 1.96$, $A_{\textrm{AGN1}} = 0.45$ and $A_{\textrm{AGN2}} = 1.15$. The cumulative reduced chi-squared associated to \ATLAS\ is $\tilde{\chi}^{2}_{\textrm{LTGs+ETGs}} = 3.47$, while the one associated to MaNGA DynPop is higher, $\tilde{\chi}^{2}_{\textrm{LTGs+ETGs}} = 5.35$, implying a lower agreement between MaNGA DynPop and this best-fit simulation.

By repeating the bootstrap procedure detailed in Section \ref{sec:ranking_method}, including also SPARC, we obtained the constraints reported in the first three rows of Table \ref{tab:bootstrap_results_LTGs_plus_ETGs}. The constraints considering the 16th, 50th and 84th percentile taken directly from the empirical CDFs instead are reported in Appendix \ref{tab:bootstrap_results_LTGs_plus_ETGs_1_sigma}. The heat maps showing the regions of lowest $\tilde{\chi}^{2}$ are, instead, described and shown in Appendix \ref{sec:correlations_parameters_heatmaps}. An assessment of the effects of a potential selection bias due to the fact that there is a higher number of low-mass galaxies with respect to high-mass galaxies is, finally, discussed in Appendix \ref{sec:selection_bias}.

Regarding the cosmological parameters, adding the comparison between simulated LTGs and SPARC to the analysis does not change significantly the parameters for SPIDER, while for \ATLAS\ there is a slight increase in the median value of $\sigma_{8}$ and $S_{8}$. The most significant impact is observed on the MaNGA DynPop constraints, where adjustments in $\Omega_{\textrm{m}}$ and $\sigma_{8}$ have resulted in $S_{8}$ being aligned with the findings of the \cite{DESandKiDS2023}, albeit incompatible with those of the \cite{Planck2018}. Overall, in all three scenarios there is now a tendency towards lower values of $S_{8}$. This is compatible with the fact that, in the $S_{8}$ tension framework, probes of the local universe show predictions of $S_{8}$ which are systematically lower than early-epoch investigations (see \citealt{Abdalla2022}, section 5).

Regarding the astrophysical parameters, we find that adding the comparison between simulated LTGs and SPARC to the analysis results in a reduction of SPIDER's constraint on $A_{\textrm{SN1}}$ below the fiducial unit value, with all three values compatible within the uncertainties. On the other hand, the effects on $A_{\textrm{SN2}}$, $A_{\textrm{AGN1}}$ and $A_{\textrm{AGN2}}$ are negligible.

Fig. \ref{fig:LTG_plus_ETG_best_fit_trends} shows the comparison between both LTGs and ETGs from these simulations and the observed trends. While there is a general agreement between observed and simulated galaxies, the left column reveals a high scatter associated to the simulated LTGs in the size-mass relation. We also notice a systematic overestimate of stellar half-mass radii at high stellar mass. Overall, the agreement is poorer with MaNGA DynPop than with \ATLAS.

Summing up, we do not find the best-fit simulation of Paper I even when constraining both LTGs and ETGs. Instead, we find that a value of $S_{8}$ generally compatible with local Universe results such as \cite{DESandKiDS2023}, and for \ATLAS\ and MaNGA DynPop a common best-fit simulation, with more reasonable cosmological parameters for the latter. In all cases, we have a low value for $A_{\textrm{SN1}}$, recovering the main result from Paper I for all three observational datasets.

\begin{figure}
    \centering
    \includegraphics[width=1\columnwidth]{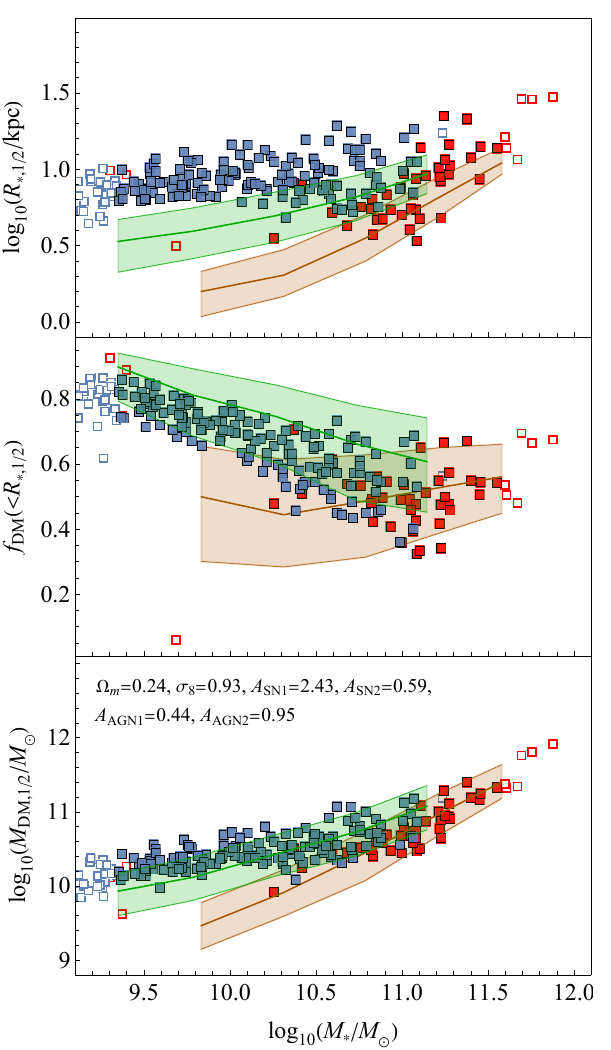}
    \caption{Same as Fig. \ref{fig:fiducial_ETG_trend}, but for the best-fit simulation to the whole MaNGA DynPop sample, `LH\_531'. The blue squares indicate simulated LTG galaxies, while red squares indicate simulated ETG galaxies. The open squares are galaxies not used for the evaluation of the cumulative reduced chi-squared associated to the respective observational trend. The dark orange region is associated to the observed MaNGA DynPop ETG sample, while the green region is associated to the LTG sample.}
    \label{fig:MANGA_LTG_plus_ETG_best_fit_trends}
\end{figure}
\begin{figure*}
    \centering
    \includegraphics[width=\textwidth]{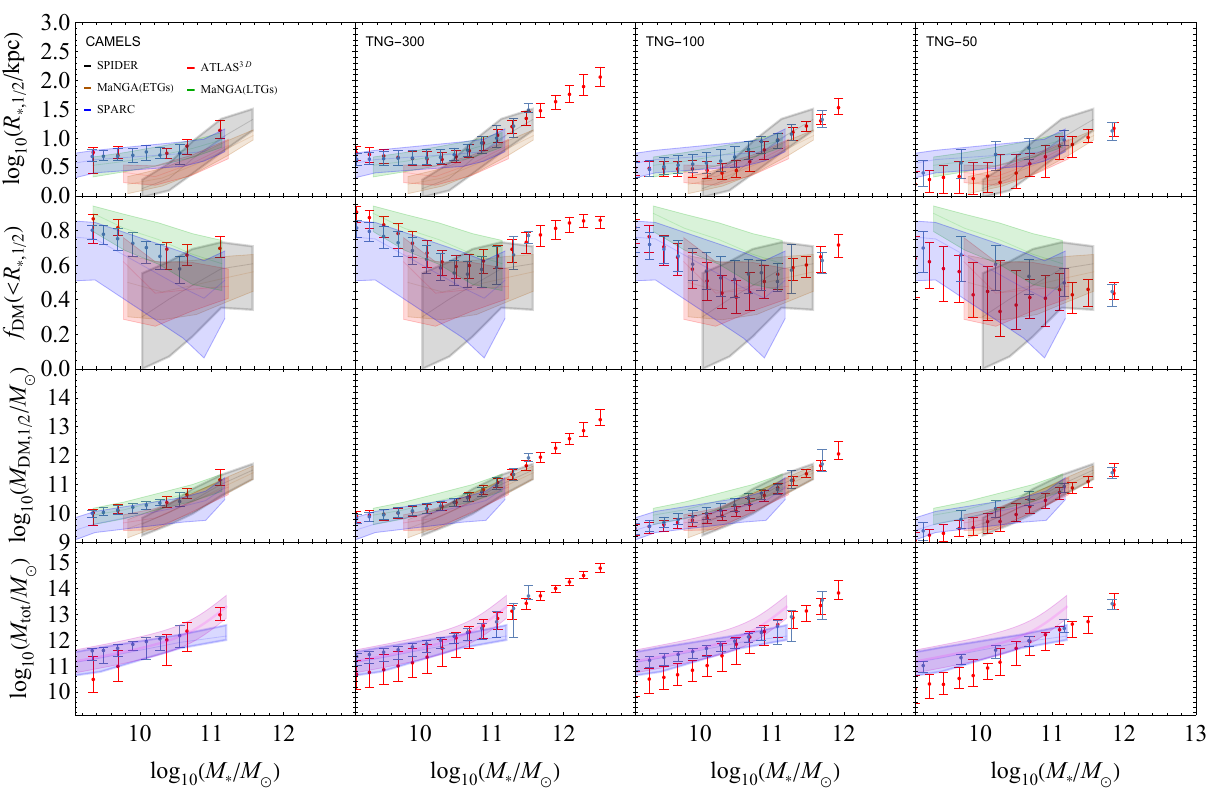}
    \caption{From top to bottom: stellar half-mass radius, DM fraction within the stellar half-mass radius and DM mass within the stellar half-mass radius as a function of stellar mass, for fiducial CAMELS simulation (first column), TNG-300 (second column), TNG-100 (third column) and TNG-50 (fourth column). The simulated galaxies for all four simulations have been binned in bins of stellar mass (red points for ETGs, blue points for LTGs, uncertainty given by the standard deviation of the y-values in the bins). The colored regions have the same meaning as in Fig.s \ref{fig:ETG_astrophysical_parameters_variation} and \ref{fig:LTG_plus_ETG_best_fit_trends}.}
    \label{fig:original_TNG_comparison_trends}
\end{figure*}
In a second analysis, we tried to constrain both LTGs and ETGs associated to the MaNGA DynPop sample. This is important because, differently from the analysis described so far, in this case the two sub-samples come from the same sample, and thus rely on the same data analysis and modelling assumptions. Using the same procedure we used in this section, we found that the best-fit simulation is the simulation `LH\_531', with the following cosmological parameters: $\Omega_{\textrm{m}} = 0.24$, $\sigma_{8} = 0.93$, $S_{8} = 0.84$, $A_{\textrm{SN1}} = 2.43$, $A_{\textrm{SN2}} = 0.59$, $A_{\textrm{AGN1}} = 0.44$ and $A_{\textrm{AGN2}} = 0.95$, with a cumulative reduced chi-squared of $\tilde{\chi^{2}}_{\textrm{LTGs+ETGs}} = 8.77$, indicating a rather poor fit. From Fig. \ref{fig:MANGA_LTG_plus_ETG_best_fit_trends}, a clear distinction emerges: the observed ETG sample trends have a markedly different slope and normalisation with respect to the corresponding LTG trends, a dichotomy which is not observed in any simulated galaxy sample. The best-fit simulation, aimed at simultaneously reproducing both trends, notably lacks simulated ETGs within the orange region for stellar masses lower than $10^{10.6}\,M_{\odot}$. This could imply a failure in the \textsc{camels} IllustrisTNG simulations in reproducing dichotomic trends for scaling relations that span a wider range of galaxy types. 

By repeating the bootstrap procedure detailed in Section \ref{sec:ranking_method}, comparing both the LTG and ETG observed trends from MaNGA DynPop with the respective types of simulated galaxies, we obtained the constraints reported in the fourth row of Table \ref{tab:bootstrap_results_LTGs_plus_ETGs}. Constraints for $S_{8}$ are compatible with both \cite{Planck2018} and \cite{DESandKiDS2023} results within $1\sigma$. Constraints for $A_{\textrm{SN1}}$ show a very high uncertainty, possibly as a consequence of the aforementioned dichotomy. The constraint for $A_{\textrm{SN1}}$ is also higher than the fiducial value, but this seems to be mainly an effect of neglecting the contribution of the gas mass for the observed galaxies: indeed, taking it into account in the analysis as described in Appendix \ref{sec:observational_biases}, we obtain a result consistent with the constraints using SPARC as the observational LTG sample.

\subsection{Comparison with original IllustrisTNG simulations}\label{sec:CAMELS_TNG_comparison}
Another possibility for the shifts found among data and simulations, which we did not account for in Paper I, is the fact that there could be mass and/or volume resolution effects at play that skew the data points upwards with respect to a `converged' simulation. 

To investigate any potential systematic shifts between observations and the fiducial \textsc{camels} simulation that may arise from convergence issues due to a low mass and volume resolution of \textsc{camels}, we have also analysed the same scaling relations used in Section \ref{sec:results}, but using the subhalos from TNG-300, TNG-100 and TNG-50. Fig. \ref{fig:original_TNG_comparison_trends} shows a comparison between the trends of the fiducial \textsc{camels} simulation and those of the three TNG simulations: in all four cases, the galaxies have been binned in bins of stellar mass, with uncertainties along the y-axis evaluated as the standard deviation of the corresponding values in each bin, in order to capture the general trend of the simulations, and to compare them to the observed trends.

As one can see from the figure, for ETGs \textsc{camels} and TNG-300 overestimate the observed trends, especially on the low-mass end of the observed trends; TNG-100, on the other hand, is much closer to the median observed trends. TNG-50, instead, seems to slightly underestimate the observed trends, especially SPIDER. Moreover, for LTGs, we can see an overestimate of \textsc{camels} and TNG-300 with respect to the SPARC observational trends, with TNG-100 and TNG-50 better reproducing them, but we can also observe that \textsc{camels} and TNG-300 reproduce the $R_{*,1/2}$-$M_{*}$ and $M_{\textrm{DM},1/2}$-$M_{*}$ trends of MaNGA DynPop for the LTG sample better than or equally good than TNG-100 and TNG-50. In particular, both TNG-100 and TNG-50 underestimate the MaNGA DynPop (LTGs) $M_{\textrm{DM},1/2}$-$M_{*}$ trend, while TNG-300 and \textsc{camels} reproduce it correctly. Moreover, in both \textsc{camels} and the three original TNG simulations, we can see that the dichotomic trend of the MaNGA DynPop sample is not reproduced; rather, both LTGs and ETGs tend to follow roughly the same trend.

Finally, to ensure that these effects are indeed due to the mass resolution rather than to other factors, e.g. the different volumes in the three TNG simulations, we compared the three simulations TNG100-1, TNG100-2 and TNG100-3, which have the same fixed volume but different mass resolutions. In particular, TNG100-2 has roughly the same mass resolution as TNG300-1, as can be seen from Table \ref{tab:IllustrisTNG_resolution_table}. Results show that the effects of changing resolution at fixed volume are the same of those shown in Fig. \ref{fig:original_TNG_comparison_trends}, that is, lower resolutions correspond to a systematic increase in DM fraction within the stellar half-mass radius.

These results seem to imply that:

\begin{enumerate}
    \item even accounting for convergence effects related to the mass and/or the spatial resolution of the \textsc{camels} fiducial simulation, the higher-resolution simulations still fail to fully replicate all observed trends. On the contrary, in some instances lower-resolution simulations perform better in matching observed trends than their higher-resolution counterparts, as in the case of the LTG sample of MaNGA DynPop;
    \item IllustrisTNG, similarly to \textsc{camels} simulations, fails to reproduce observed dichotomic trends between LTGs and ETGs. The subhalos are aligned at all resolutions, and roughly follow the same trend, differently from the observed scaling relation trends. The fact that \textsc{camels} and TNG-300 do not manage to reproduce the MaNGA DynPop observed trend for ETGs could be a consequence of this point.

\end{enumerate}

\subsection{Comparison with literature}
In the past, studying the impact of the variation of cosmological and astrophysical parameters on the physics of galaxies was very hard, given the computational cost of running multiple hydrodynamical simulations. There have been, however, some initial attempts to produce simulations with differing values of astrophysical parameters and study the effect that their variation have on galaxies. One of such attempts was the SEAGLE programme \citep{Mukherjee2018,Mukherjee2021,Mukherjee2022}, which used simulated strong gravitational lenses from the EAGLE simulations to study galaxy formation. In particular, \cite{Mukherjee2021} explore the impact of SN- and AGN-feedback parameters on early-type deflectors' observables. They find that a low stellar feedback better matches the size-mass relation derived from the Sloan Lens ACS Survey (SLACS) observations, which is consistent with our findings. However, they also find that different AGN feedback parameters, such as the viscosity parameter or the AGN heating temperature, have an important effect on the simulations. Given that the kinetic AGN feedback parameters $A_{\textrm{AGN1}}$ and $A_{\textrm{AGN2}}$ seem to have a negligible impact on simulations, in future works we will analyse the new \textsc{camels} simulations, that vary more AGN feedback parameters, which could be more impactful on the scaling relation trends (Tortora et al. in prep.).

With better computational power and simulation techniques, simulation suites such as \textsc{camels} became available. There have been many works in literature that used these simulations to infer the values of cosmological parameters, for example by using photometry of galaxies \citep{Hahn2024}, the physical properties of single galaxies \citep{Villaescusa-Navarro2022_one_galaxy} or of multiple galaxies \citep{Chawak2024}. Using the photometry of multiple galaxies, \cite{Hahn2024} manage to constrain with remarkable accuracy the values of $\Omega_{\textrm{m}}$ and $\sigma_{8}$, while the work of \cite{Chawak2024} shows that, by training a neural network to obtain the cosmological and astrophysical parameters of \textsc{camels} simulations given the properties of two galaxies, the recovery of $\sigma_{8}$ and $A_{\textrm{SN2}}$ is weak (with the recovery getting better by using ten galaxies instead of two), while AGN-feedback parameters are completely unconstrained. These results are also consistent with our findings.

Other attempts in literature to constrain cosmological and astrophysical parameters with the use of simulations include HIFLOW \citep{Hassan2022}, a generative model for the estimation of the probability density function of the target observable (e.g. cosmological parameters), based on the masked autoregressive flow method \citep{Papamakarios2017}. The advantage with respect to our method is that HIFLOW manages to find the correct posterior distribution of the cosmological parameters $\Omega_{\textrm{m}}$ and $\sigma_{8}$ given the observed \textsc{Hi} maps from \textsc{camels} and uniform priors. A disadvantage is that the method is weakly dependant on the value of the astrophysical parameters.

Another approach is to train a neural network to work as an emulator of the \textsc{camels} simulations, to perform fast implicit likelihood inference (ILI, \citealt{Jo2023}) of the cosmological and astrophysical parameters by substituting the simulations with the emulators. By using the cosmic star formation rate density (SFRD) and the stellar mass functions (SMF), the method is able to obtain the posterior distributions for $\Omega_{\textrm{m}}$, $\sigma_{8}$ and the SN- and AGN-feedback parameters. By applying this procedure on actual observational data from \cite{Leja2020,Leja2022}, \cite{Jo2023} find very extreme values for the cosmological parameters $\Omega_{\textrm{m}}$ and $\sigma_{8}$, which they report is due to \textsc{camels} resolution effects and degeneracies with the astrophysical parameters. This issue, however, seems to not be present in our constraining procedure. They also notice how AGN-feedback parameters $A_{\textrm{AGN1}}$ and $A_{\textrm{AGN2}}$ are weakly constrained with their procedure, due to a low influence of kinetic AGN feedback on formation of high stellar mass galaxies and star formation in massive galaxies. This is in agreement with our results.

There exist many other projects that try to make simulation-based inference of cosmological and astrophysical parameters, for example by using graph neural networks to obtain information from galaxy distributions \citep{Roncoli2023}, but an extensive review of these is outside the scope of our work.

\section{Conclusions}\label{sec:conclusions}
In this work, we expanded the \textit{CASCO} project \citep{Busillo2023}, by extending the analysis to early-type galaxies. This time, we considered three different observational datasets (SPIDER, \ATLAS\ and MaNGA DynPop) and updated both the definition and the evaluation procedure of the reduced chi-squared, by considering a selection of only those simulations which have an uniform coverage in stellar mass. Our main results are the following:

\begin{itemize}
    \item We have shown that both the fiducial simulation, which is the one having the cosmological and astrophysical parameters equal to the original IllustrisTNG simulation, and the best-fit simulation of Paper I that we found for LTGs are not a good fit for the simulated early-type galaxies. Both simulations systematically overestimate all three observational trends. Even when constraining both LTGs and ETGs at the same time, we do not recover the two simulations as the best-fit ones.
    \item We ran our procedure (see Section \ref{sec:ranking_method}), searching for the best-fit between the simulated ETGs and the three different observational trends. We found for the SPIDER sample that the best-fit simulation is the simulation `LH\_523', with constraints that show values for $A_{\textrm{SN1}}$ and $A_{\textrm{SN2}}$ higher and lower than the fiducial unit values from \textsc{camels}, respectively. This is incompatible with the results obtained in Paper I from the SPARC observational sample, which showed $A_{\textrm{SN1}}$ lower and $A_{\textrm{SN2}}$ higher than the respective fiducial values. For the \ATLAS and MaNGA DynPop samples, we instead found values of the astrophysical parameters in line with the results from Paper I, although the latter shows respectively lower and higher values of $\Omega_{\textrm{m}}$ and $\sigma_{8}$ than those from \cite{Planck2018} and \cite{DESandKiDS2023}, resulting in a very low value of $S_{8}$. Constraints from the ranking procedure are thus strongly affected by the properties of the reference observational trends. Similarly to Paper I, we find also in this case that with all observational samples the AGN-feedback parameters $A_{\textrm{AGN1}}$ and $A_{\textrm{AGN2}}$ are unconstrained.
    \item Constraints are modified when considering also LTG observations along with ETG observations. Using SPARC as the observational LTG sample and performing the bootstrap procedure, we obtain a lower value of $S_{8}$ with respect to \cite{Planck2018} and a lower value of $A_{\textrm{SN1}}$ with respect to the fiducial \textsc{camels} unit value for all three ETG observational datasets, recovering the main result of Paper I. Using the full MaNGA DynPop sample (including both ETGs and LTGs) results in an $S_{8}$ value compatible with \cite{DESandKiDS2023} and a very large uncertainty on $A_{\textrm{SN1}}$, possibly due to a dichotomic trend between the ETGs and the LTGs in the MaNGA DynPop sample.
    \item To check for a possible systematic effect due to the limited numerical resolution in \textsc{camels}, we compared the fiducial simulation with the original Illustris simulations. We have considered ETGs and LTGs from TNG-300, TNG-100 and TNG-50, and showed that both the fiducial \textsc{camels} simulation and TNG-300 are compatible with each other as far as the behavior with respect to the observational trends is concerned, with both simulations overestimating the observed trends in a similar way, both for LTGs and ETGs.
Results suggest that systematic effects associated to both simulations and observational datasets are influencing the alignment of the respective trends. In particular a lack of convergence caused by a low particle mass resolution does not imply that simulations with low resolution behave necessarily worse than higher resolution ones.
\end{itemize}

In this work, we have seen that there are many caveats to be considered before fully exploiting the predictive power of galaxy scaling relations to constrain cosmology and astrophysics. Expanding our scope to early-type galaxies has shown that fitting only one type of galaxy to the observational datasets does not imply that the fit generalizes to other types of galaxies, and we have shown that, even by changing the reference observational trends, the constraints show a significant variation. 

We have also shown that, despite the low mass and volume resolution of \textsc{camels}, the fiducial simulation is able to reproduce the MaNGA DynPop observational trend for LTGs. The comparison with the original TNG simulations shows that, in regards to reproducing observed scaling relations, the IllustrisTNG suite of \textsc{camels} is still a reliable tool, even with the presence of possible convergence effects. 
Both \textsc{camels} and the original IllustrisTNG simulations, however, fail in reproducing the dichotomic trend shown by the full MaNGA DynPop observational sample. This could imply that there are some limitations in how subhalos properties are obtained in the simulations (and as such, showing some of the limits of the sub-grid approach used to replicate the baryonic processes in the IllustrisTNG subhalos), or in the various observational samples used for the comparisons, given that different approaches are used to obtain them.

For the future, we plan to analyse the evolution of simulated scaling relations across cosmic time \citep{Tortora+14_DMevol, Tortora+18_KiDS_DMevol, Sharma2022}, to check whether the simulations manage to reproduce correctly the observations also at high redshift. Constraints of scaling relations at high redshift could also provide some information about past values of $\Omega_{\textrm{m}}$ and a possible evolution of $\sigma_{8}$ with cosmic time \citep{Adil2024}. Moreover, in this work we fixed for simplicity the sub-grid physics to the one from IllustrisTNG. Given that no recipe is a perfect representation of the real astrophysical processes that occur in the Universe, it is important to check that the constraining procedure is robust with respect to the change of subgrid recipe. While some of this analysis has been performed for LTGs only in Paper I,  in future works we will explore more in detail the impact of changing the sub-grid physics, especially with the new \textsc{camels} simulations, which provide new suites such as Magneticum and EAGLE.

Other simulations from \textsc{camels}, with higher resolution and larger volumes, could also allow us to check further for convergence effects. In future works, especially by leveraging on the enhanced statistics at high mass by using the simulations having larger volumes, it may be possible to check the effects that additional AGN-feedback parameters present in the new \textsc{camels} simulations \citep{Ni2023}, such as the normalisation factor for the Bondi rate for the accretion onto SMBHs or the threshold between the low-accretion and high-accretion states of AGN feedback \citep{Weinberger2017}, have on the scaling relations, and in particular on the transition mass scale for scaling relations such as the total-stellar mass relation, the so-called `golden mass' (Tortora et al. in prep.).

Future surveys, like the Euclid Wide Survey and the Cosmic Dawn Survey \citep{EuclidWide2022,EuclidDawn2022}, will both detect much larger samples of rare, massive galaxies and observe galaxies up to very high redshifts. In particular, it is expected that around $10^{5}$ strong gravitational lenses will be found by Euclid \citep{Collett2015}, enabling the determination of very precise mass estimates and constraints on dark matter fraction for a substantial set of high-mass galaxies. This dataset can then be utilised for comparison with simulations. As such it is auspicable that \textsc{camels} will allow in the future to have more statistics available, especially on the high-mass end of simulated galaxies.

\begin{acknowledgements}
V.B., C.T. and M.S. acknowledge the INAF grant 2022 LEMON.
\end{acknowledgements}

\bibliography{bibliography}

\begin{appendix}
\section{Observational biases}\label{sec:observational_biases}
The simulated datasets are based on quantities obtained from the \textsc{subfind} algorithm, which are 3D quantities that are not affected by projection effects. On the contrary, the SPIDER, \ATLAS and MaNGA DynPop observational datasets are based on inferred physical properties, such as the effective radius, which are subject to both observational effects and model assumptions for the galaxies. This difference could lead to biases in the inference of the cosmological and astrophysical parameters, by performing a comparison between quantities which are not exactly comparable.

While the total quantities are not affected by projection effects, galaxy sizes and the dark matter mass within a certain radius are affected by them. Regarding the sizes, the main issues are the presence of a mass-to-light ratio gradient, the fact that the effective radius is typically a projected, 2D quantity, and finally the fact that, in simulations, one takes the radius of a sphere containing half the total mass of the galaxy, without any model assumption, while in observed galaxies one fits to the data a surface brightness model (e.g. elliptical Sérsic profile) or evaluate the growth curve, determining the half-light radius as the radius at which the growth curve reaches half the total luminosity. Regarding the first issue, an $M/L$ gradient has the consequence that the half-light radius is different from the corresponding half-mass radius. Moreover, there is a marked difference between effective radii for the same galaxies measured in different optical bands, with the redder bands, such as the $r$ and $K$ bands (the latter of which is used to measure the effective radius in SPIDER, while the former is used for \ATLAS and MaNGA DynPop), having the value of the effective radius closer to the respective half-mass radius than the bluer ones, such as the $g$ band \citep{Vulcani2014,Baes2024}. It has to be noted, however, that $M/L$ gradients can have a small impact on the size-mass relation of galaxies at low redshift, of the order of $R_{\textrm{m}}/R_{\textrm{l}} \sim 0.6$, where $R_{\textrm{m}}$ is the scale which contains half the projected mass and $R_{\textrm{l}}$ the one which contains half the projected light (see e.g. \citealt{Bernardi2023, Wu2024, Baes2024}). The projection effects also have a small impact, given that $R_{*,1/2}/R_{\textrm{e}} \approxeq 4/3$ is accurate to better than 2 per cent for most surface brightness profiles \citep[Appendix B]{Wolf2010}. Finally, the assumption of an elliptical light profile should have a negligible impact with respect to assuming a spherical light profile like we have done with the SPIDER and \ATLAS datasets (e.g. Singular Isothermal Sphere, \citealt{Tortora2012}). While surely the model assumption produces some biases with respect to evaluating the half-mass radius directly from the simulation particles, the precise quantification of this bias is outside the scope of this work.

Regarding the DM mass within the half-mass radius, there are two main sources of biases. First, the assumption of a mass model for the galaxies, either assuming a cumulative model for DM plus baryons (like in the SPIDER and \ATLAS observational samples) or a model with two mass components for DM and baryons (e.g. generalized Navarro-Frenk-White profile for DM in MaNGA DynPop) can affect the total and DM mass values. Second, neglecting the gas mass in observations can influence directly the DM mass within the half-mass radius, given that $M_{\textrm{DM},1/2} = M_{\textrm{tot},1/2} - M_{*,1/2} - M_{\textrm{gas},1/2}$. Regarding the former, we have checked for MaNGA DynPop that changing the assumption of the underlying dark matter model minimally influences the observational datasets, with the scatter between the medians for all scaling relation trends of all the different DM models being at most $0.02\,\textrm{dex}$. Regarding the latter, for ETGs such as those in the SPIDER and \ATLAS datasets, neglecting the gas contribution should not influence strongly the observational values of $M_{\textrm{DM},1/2}$, given that the amount of gas in the central regions of these galaxies is negligible. In the SPARC LTG sample, instead, the mass of the gas is taken into account, so there is no bias in that sample. Such an issue is however present for the MaNGA DynPop LTG sample, given that the gas mass is not reported in their catalog. To check for possible systematic effects due to having the gas mass included in the DM mass in the observational MaNGA DynPop sample, we searched for the best-fit simulation by considering for the simulations the quantities $\tilde{f}_{\textrm{DM},1/2} = (M_{\textrm{DM},1/2} + M_{\textrm{gas},1/2})/M_{\textrm{tot},1/2}$ and $\tilde{M}_{\textrm{DM},1/2} = M_{\textrm{DM},1/2} + M_{\textrm{gas},1/2}$, instead of $f_{\textrm{DM},1/2}$ and $M_{\textrm{DM},1/2}$, where $M_{\textrm{gas},1/2}$ is the sum of all the gas particles' mass within the stellar half-mass radius, as obtained via \textsc{subfind}.

\begin{figure}
    \centering
    \includegraphics[width=1\linewidth]{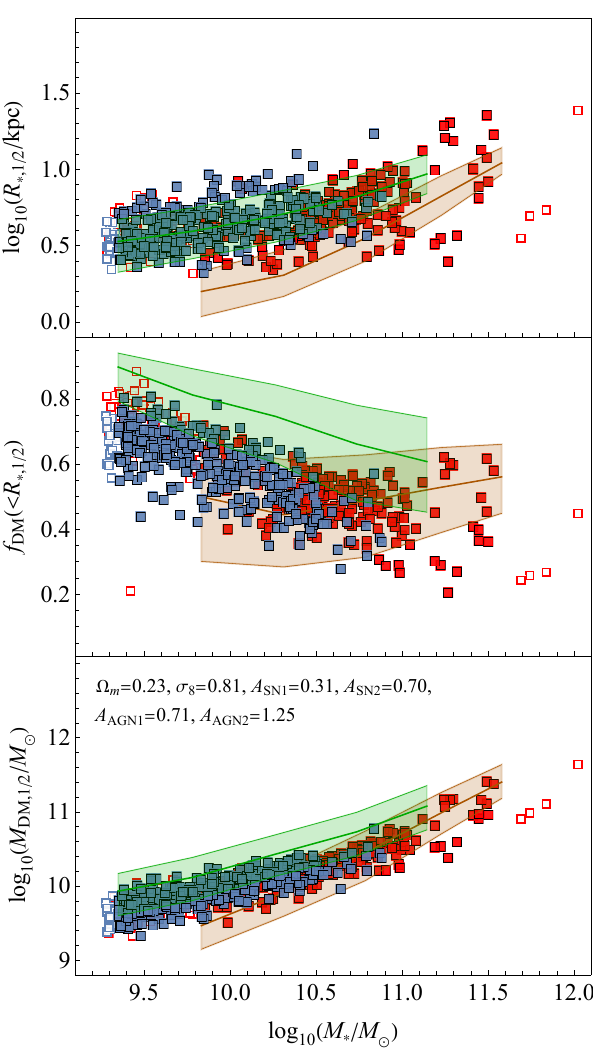}
    \caption{Same as Fig. \protect\ref{fig:MANGA_LTG_plus_ETG_best_fit_trends}, but for the best-fit simulation `LH\_717', obtained by considering the gas mass as a systematic contribution to the simulations' DM mass.}    \label{fig:MANGA_LTG_plus_ETG_best_fit_trends_alternative}
\end{figure}

We find that the new best-fit simulation is the simulation `LH\_717', with the following parameters: $\Omega_{\textrm{m}} = 0.23$, $\sigma_{8} = 0.81$, $S_{8} = 0.71$, $A_{\textrm{SN1}} = 0.31$, $A_{\textrm{SN2}} = 0.70$, $A_{\textrm{AGN1}}= 0.71$ and $A_{\textrm{AGN2}} = 1.25$, with cumulative reduced chi-squared equal to $\tilde{\chi}^{2} = 7.29$. The cosmological parameters for this simulation are within $1\sigma$ with respect to the results of Table \ref{tab:bootstrap_results_LTGs_plus_ETGs}, while the SN feedback parameters are not. In particular, $A_{\textrm{SN1}}$ in this case is lower than the fiducial unit value, a result which is in line with the constraints from SPARC + SPIDER and SPARC + \ATLAS. This simulation, however, still fails to solve the dichotomy between the LTG and ETG trends of MaNGA DynPop, as shown in Fig. \ref{fig:MANGA_LTG_plus_ETG_best_fit_trends_alternative}.

A possible way to reduce all these issues is to forward-model the simulated quantities to the observational space \citep{Dickey2021}. A full forward-modeling of the simulated quantities is, however, outside of the scope of this work.

\section{Raw parameter constraints}
In this section, we show the tables reporting the constraints obtained from the bootstrap procedures without the smoothing of the CDFs. 
\begin{table*}
\centering
\renewcommand*{\arraystretch}{1.50}
\caption{Constraints on cosmological and astrophysical parameters, obtained by bootstrapping both observational and simulated datasets and taking for each resampling the best-fit simulation.}
\label{tab:bootstrap_results_1_sigma}
    \begin{tabular}{cccccccc}
    \hline
    \hline
         Obs. Trend&  $\Omega_{\textrm{m}}$&  $\sigma_{8}$&  $S_{8}$&  $A_{\textrm{SN1}}$&  $A_{\textrm{SN2}}$&  $A_{\textrm{AGN1}}$& $A_{\textrm{AGN2}}$\\
    \hline
         SPIDER&  $0.25_{-0.05}^{+0.02}$&  $0.77_{-0.13}^{+0.13}$&  $0.67_{-0.09}^{+0.15}$&  $1.83_{-0.63}^{+0.74}$&  $0.64_{-0.11}^{+0.18}$&  $0.77_{-0.40}^{+1.91}$& $1.17_{-0.58}^{+0.25}$\\
         \ATLAS&  $0.21_{-0.03}^{+0.01}$&  $0.89_{-0.24}^{+0.04}$&  $0.72_{-0.14}^{+0.05}$&  $0.30_{-0.03}^{+0.15}$&  $1.52_{-0.47}^{+0.44}$&  $0.97_{-0.51}^{+1.00}$& $1.12_{-0.55}^{+0.32}$\\
         MaNGA DynPop&  $0.16_{-0.00}^{+0.04}$&  $0.97_{-0.17}^{+0.00}$&  $0.70_{-0.07}^{+0.00}$&  $0.30_{-0.04}^{+0.01}$&  $1.61_{-0.09}^{+0.16}$&  $2.71_{-1.91}^{+0.00}$& $1.77_{-1.16}^{+0.00}$\\
    \hline
    \end{tabular}
\tablefoot{The values reported are the 16th, 50th (median) and 84th percentiles, taken from the respective empirical CDFs, without any smoothing.}
\end{table*}

\begin{table*}
\centering
\renewcommand*{\arraystretch}{1.50}
\caption{Constraints on cosmological and astrophysical parameters, obtained by bootstrapping the observational (including SPARC) and simulated datasets, and taking for each resampling the best-fit simulation.}
\label{tab:bootstrap_results_LTGs_plus_ETGs_1_sigma}
    \begin{tabular}{cccccccc}
    \hline
    \hline
         Obs. Trend&  $\Omega_{\textrm{m}}$&  $\sigma_{8}$&  $S_{8}$&  $A_{\textrm{SN1}}$&  $A_{\textrm{SN2}}$&  $A_{\textrm{AGN1}}$& $A_{\textrm{AGN2}}$\\
    \hline
         SPARC + SPIDER&  $0.24_{-0.03}^{+0.01}$&  $0.76_{-0.15}^{+0.07}$&  $0.68_{-0.13}^{+0.04}$&  $0.70_{-0.39}^{+0.05}$&  $0.83_{-0.13}^{+0.05}$&  $0.84_{-0.46}^{+1.24}$& $1.02_{-0.34}^{+0.23}$\\
         SPARC + \ATLAS&  $0.20_{-0.00}^{+0.02}$&  $0.97_{-0.08}^{+0.00}$&  $ 0.77_{-0.03}^{+0.02}$&  $0.32_{-0.03}^{+0.07}$&  $1.09_{-0.00}^{+0.87}$&  $1.35_{-0.89}^{+0.07}$& $0.65_{-0.00}^{+0.50}$\\
         SPARC + MaNGA DynPop&  $0.22_{-0.02}^{+0.00}$&  $0.89_{-0.09}^{+0.08}$&  $0.77_{-0.06}^{+0.02}$&  $0.31_{-0.02}^{+0.08}$&  $1.41_{-0.33}^{+0.54}$&  $1.29_{-0.84}^{+0.67}$& $1.15_{-0.53}^{+0.29}$\\
    \hline
    \end{tabular}
\tablefoot{The values reported are the 16th, 50th (median) and 84th percentiles, taken from the respective empirical CDFs, without any smoothing.}
\end{table*}

\section{Chi-squared heat maps}\label{sec:correlations_parameters_heatmaps}
\begin{figure*}
    \centering
    \includegraphics[width=0.90\textwidth]{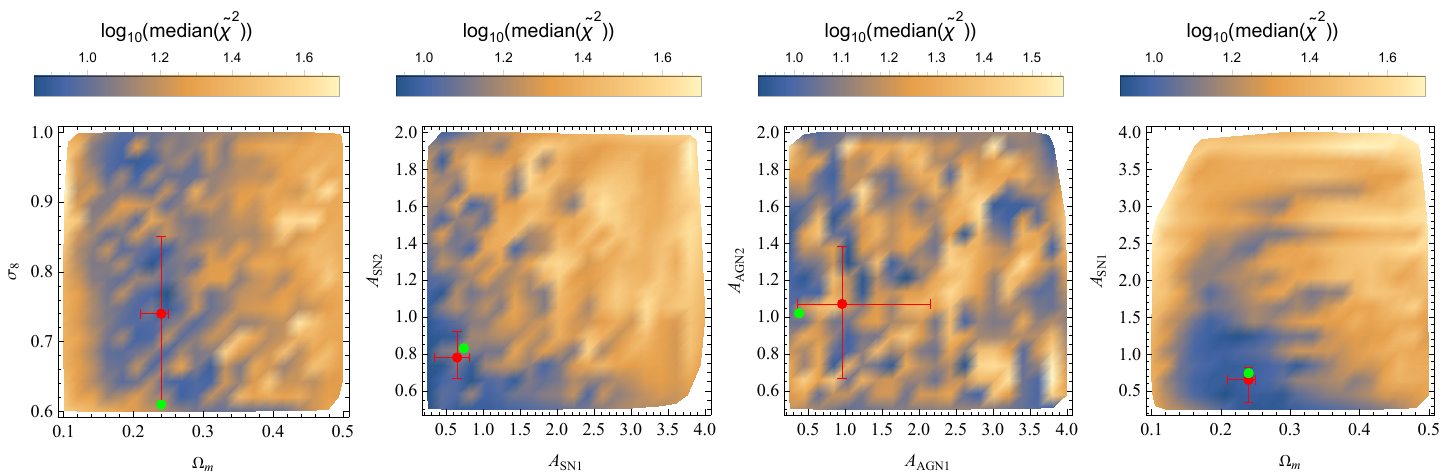}
    \includegraphics[width=0.90\textwidth]{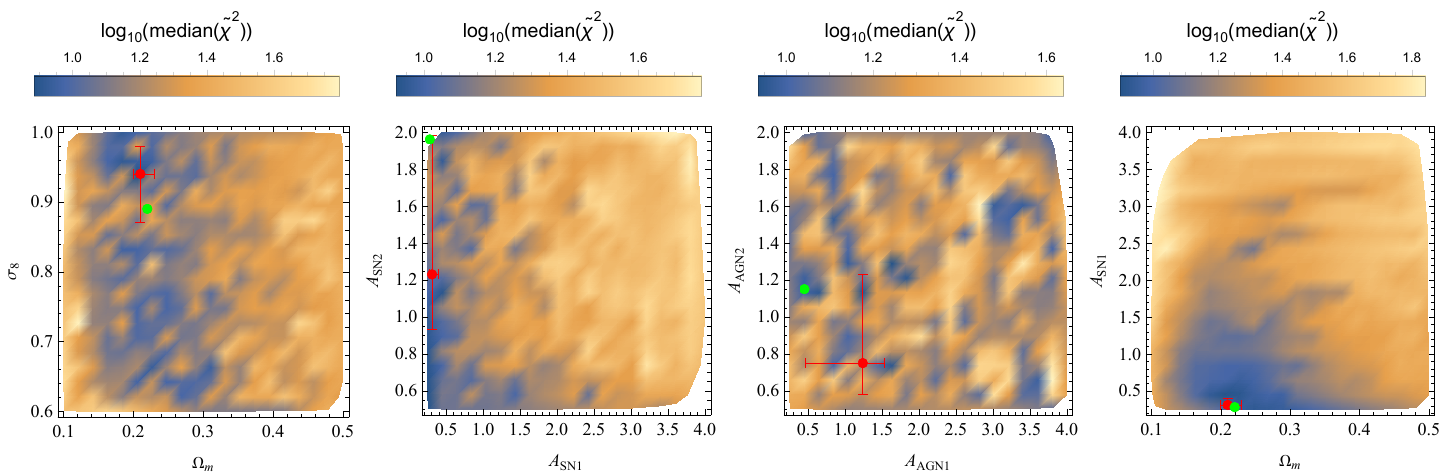}
    \includegraphics[width=0.90\textwidth]{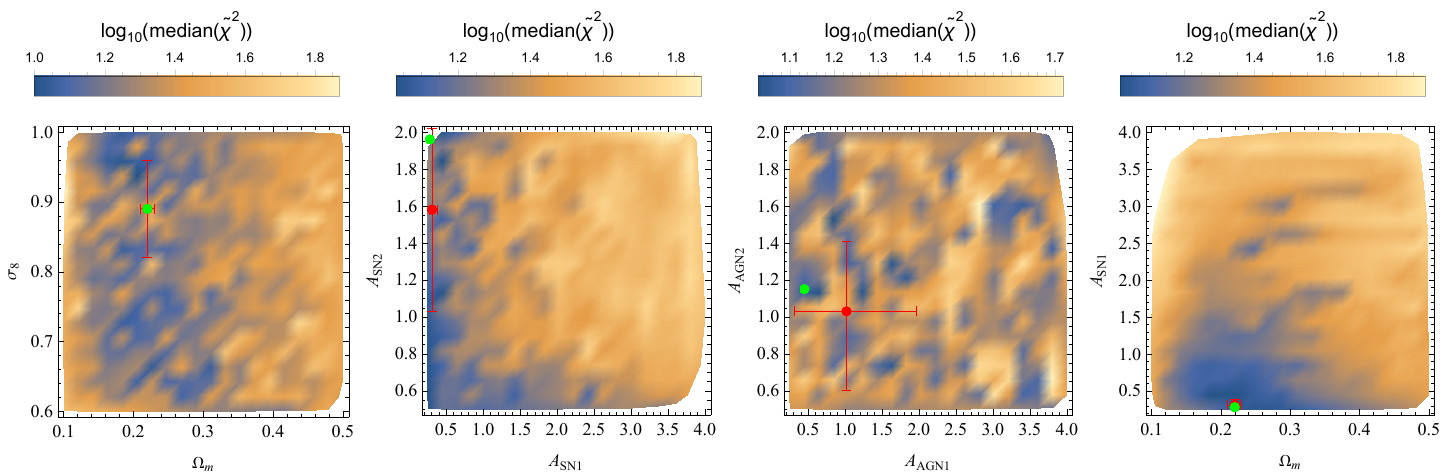}
    \includegraphics[width=0.90\textwidth]{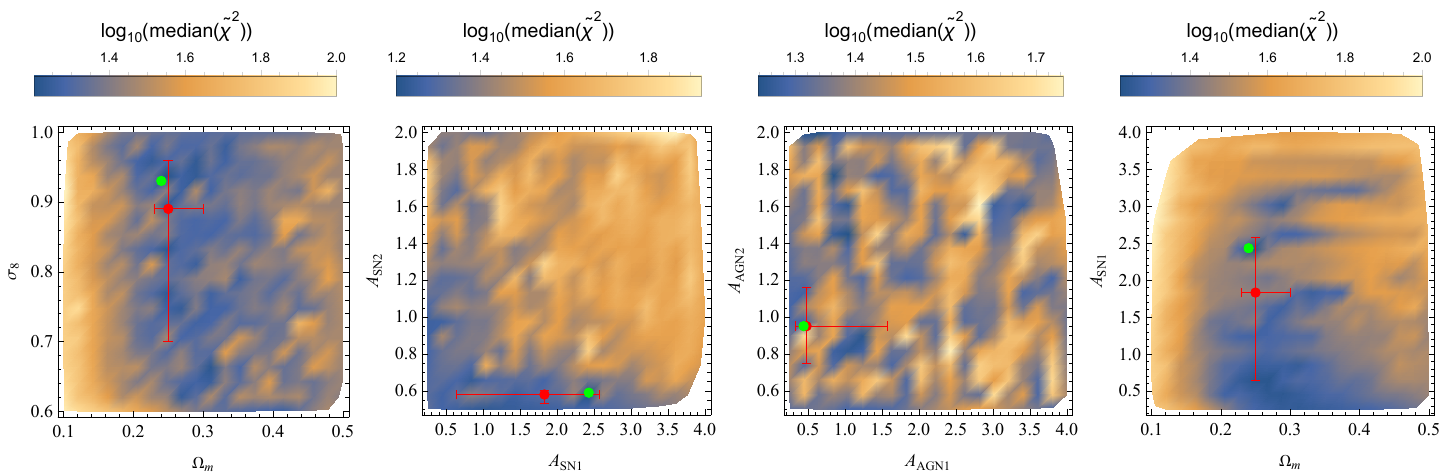}
    \caption{Heat maps showing regions of the parameter space as a function of the value of cumulative reduced chi-squared for each simulation. The colors show the logarithm of the median of the 100 values of $\tilde{\chi}^{2}$ associated to each resampling obtained by the bootstrap procedure. The rows, from top to bottom, show the heat maps for the fits with SPARC+SPIDER, SPARC+\ATLAS, SPARC+MaNGA DynPop and LTGs+ETGs from MaNGA DynPop. The red point shows the values of Table \protect\ref{tab:bootstrap_results_LTGs_plus_ETGs}, while the green points are the values from the respective best-fit simulations.}
    \label{fig:heat_map_correlations}
\end{figure*}

A potential issue in the extraction of cosmological and astrophysical parameters from the procedure described in the paper is the presence of degeneracies, which allows for simulations with wildly different parameters to have similar values of chi-squared, thus increasing the uncertainty in the parameters. It is thus useful to study the relation between different cosmological and astrophysical parameters in terms of values of chi-squared. In Fig. \ref{fig:heat_map_correlations}, we plotted heat maps that show the parameter spaces $\Omega_{\textrm{m}}$-$\sigma_{8}$, $A_{\textrm{SN1}}$-$A_{\textrm{SN2}}$, $A_{\textrm{AGN1}}$-$A_{\textrm{AGN2}}$ and $\Omega_{\textrm{m}}$-$A_{\textrm{SN1}}$. These planes give complete information about the parameter space, and thus other figures with different combinations would be redundant. The plots are color-coded with the value, for all pair of points associated to each simulation, of the logarithm of the median of the 100 values of the cumulative $\tilde{\chi}^{2}$ obtained by resampling the respective simulation. Regions of low (high) chi-squared are shown in blue (orange). Each row shows a different observational dataset from which the values of $\tilde{\chi}^{2}$ were obtained, with the same order as that of Table \ref{tab:bootstrap_results_LTGs_plus_ETGs}. We avoided reporting the plots for the ETG-only observational datasets to avoid redundancy, given that the plots are very similar to those shown in the first three rows of Fig. \ref{fig:heat_map_correlations}.

From the figure, we deduce that in all cases there is a degeneracy in the $\Omega_{\textrm{m}}$-$\sigma_{8}$ plane, which is roughly given by a vertical band. This strong degeneracy along the $\sigma_{8}$ axis seems to be associated to the uncertainty in the $\sigma_{8}$ parameter estimation. A degeneracy in the $A_{\textrm{SN1}}$-$A_{\textrm{SN2}}$ plane is also visible, with an hyperbole-shaped blue region in the lower left corner of the parameter space. This degeneracy seems to constrain either $A_{\textrm{SN1}}$ or $A_{\textrm{SN2}}$ to have low values, with the other parameter having a large uncertainty. We cannot detect any degeneracy in the $A_{\textrm{AGN1}}$-$A_{\textrm{AGN2}}$ plane, while there is a weak vertical degeneracy in the $\Omega_{\textrm{m}}$-$A_{\textrm{SN1}}$ plane, which seems to affect mainly the constraints with the MaNGA DynPop LTGs+ETGs sample. In all cases, the best-fit simulation is within $1\sigma$ of the bootstrap constraints. There are cases, however, in which the best-fit simulation is on the edge of the $1\sigma$ error bar, mostly associated with low constraining power situations, such as with $\sigma_{8}$ and both AGN feedback parameters.

\section{Assessing selection bias due to $\chi^{2}$ definition}\label{sec:selection_bias}
With the definition of equation \eqref{eq:chi_squared_new_definition}, the evaluation of the chi-squared is dominated by low-mass simulated galaxies, such as dwarf ellipticals or LTGs, which are much more numerous than high-mass galaxies. Our approach could thus be biased towards simulations that do a better job in reproducing the low-mass end of the scaling relations. To assess whether this asymmetry introduces such a selection bias, we tried to reproduce the results for IllustrisTNG in Paper I by finding the constraints for the cosmological and astrophysical parameters with respect to the observational SPARC sample, this time by eliminating a fraction of LTGs under a threshold value $M_{\textrm{thr.}} = 10^{10}\,M_{\odot}$ such that the number of LTGs below this threshold value is equal to the number of LTGs above the threshold value. Results show that the discrepancies between the values of the newly constrained parameters and the constraints obtained in Paper I are negligible. We thus conclude that this selection bias does not influence significantly the results of our work.
\end{appendix}
\end{document}